%% file: main.tex
\documentclass[letterpaper]{article} 
\usepackage{float}
\usepackage[pdftex]{graphicx}
\usepackage{fixltx2e}
\usepackage{tabularx}
\usepackage{verbatim}
\usepackage{color}
\usepackage{xparse}
\usepackage{xmpmulti}
\usepackage{transparent}
\usepackage{fancyhdr}
\usepackage{cite}
\usepackage{rotating}
\usepackage{authblk}
\usepackage{amssymb,amsmath,float}
\usepackage[linesnumbered, algoruled, vlined]{algorithm2e}
\usepackage{booktabs}
\usepackage{subcaption}
\usepackage{tikz}
\usepackage{amsthm}
\usepackage{hyperref}
\usepackage{wrapfig}
\usepackage{setspace}
\usepackage{graphicx}
\usepackage{epsfig}
\usepackage[all]{xy}
\usepackage{verbatim}
\usepackage{float}
\usepackage{bm}
\usepackage{multirow}
\usepackage[compact]{titlesec}
\usepackage{diagbox}
\usepackage{array}
\usepackage{units}
\usepackage[export]{adjustbox}
\usepackage{cleveref}
\usetikzlibrary{arrows, automata, positioning}
\tikzset{auto, >=stealth}
\tikzset{every edge/.append style={shorten >= 1pt}}
\tikzset{main node/.style={circle,draw,minimum size=1cm,inner sep=0pt}}

\newtheorem{assumption}{Assumption} 
\newtheorem{lemma}{Lemma}            

\newtheorem{problem}{Problem}
                       
\newtheorem{definition}{Definition}

\let\oldIEEEkeywords\IEEEkeywords
\def\IEEEkeywords{\oldIEEEkeywords\normalfont\bfseries\ignorespaces}
\ifdefined\beginthm\else
\newtheorem{thm}{Theorem}
\fi
\usepackage{pdfpages}

\title{Reinforcement Learning With Reward Machines in Stochastic Games}

\author[1]{Jueming Hu}
\author[2]{Jean-Rapha\"el Gaglione}
\author[1]{Yanze Wang}
\author[1,*]{Zhe Xu}
\author[2]{Ufuk Topcu}
\author[1]{Yongming Liu}

\affil[1]{Arizona State University,USA}

\affil[2]{University of Texas at Austin, USA}
\affil[*]{Corresponding author. Email: xzhe1@asu.edu}

\date{}

\usepackage{aliasing}
\usepackage{macros}

\begin{document}

\maketitle


\begin{abstract}
   We investigate multi-agent reinforcement learning for stochastic games with complex tasks, where the reward functions are non-Markovian.
    We utilize reward machines to incorporate high-level knowledge of complex tasks. 
    We develop an algorithm called \emph{Q-learning with reward machines for stochastic games ({\algoName{}})}, to learn the \emph{best-response strategy} at Nash equilibrium for each agent. 
    In {\algoName{}}, we define the \emph{Q-function at a Nash equilibrium} in \emph{augmented state space}. The augmented state space integrates the state of the stochastic game and the state of reward machines. Each agent learns the Q-functions of all agents in the system. We prove that Q-functions learned in QRM-SG converge to the Q-functions at a Nash equilibrium if the stage game at each time step during learning has a global optimum point or a saddle point, and the agents update Q-functions based on the best-response strategy at this point. 
    We use the Lemke-Howson method to derive the best-response strategy given current Q-functions.
    The three case studies show that {\algoName{}} can learn the best-response strategies effectively. {\algoName{}} learns the best-response strategies after around 7500 episodes in Case Study I, 1000 episodes in Case Study II, and 1500 episodes in Case Study III, while baseline methods such as Nash Q-learning and MADDPG fail to converge to the Nash equilibrium in all three case studies.

    
\end{abstract}

\input{introduction.tex}

\input{related_work}
\input{preliminaries}

\input{problem}

\input{solution}

\input{experiment}



\input{conclusion}

\section*{Acknowledgements}
This work is supported by NASA University Leadership Initiative program (Contract No. NNX17AJ86A, PI: Yongming Liu, Technical Officer: Anupa Bajwa). This work is also supported by grants NSF CNS 2304863 and ONR N00014-23-1-2505 (for Zhe Xu and his students), and ONR N00014-22-1-2254 and ARO W911NF-20-1-0140  (for Ufuk Topcu and his students).

\bibliographystyle{IEEEtran}
\bibliography{references}

\newpage
\input{appendix}

\end{document}

%% file: introduction.tex
\section{Introduction}
\label{sec:introduction}
%
%
%



In games, multiple agents interact with each other and the behavior of each agent can affect the performance of other agents.
\emph{Multi-agent reinforcement learning} (MARL)~\cite{Buoniu2010MARL} is a framework for multiple agents to optimize their strategies by repeatedly interacting with the environment.
The interactions between agents in MARL problems can be modeled as a \emph{stochastic game} (SG), where Nash equilibrium can be a solution concept.
In a Nash equilibrium, agents have \emph{best-response} strategies considering other agents' actions.

In many complex games, the reward functions are non-Markovian (i.e., the reward of each agent can depend on the history of events).
One way of encoding a non-Markovian reward function is by using a type of Mealy machine named \emph{reward machines}~\cite{icarte2018}.
In this paper, we focus on stochastic games where each agent intends to complete a complex task which can be represented by a reward machine capturing the temporal structure of the reward function. 
%


%

\begin{figure}[ht] 
    \centering
    
    \begin{subfigure}[t]{.40\linewidth}
        \centering
        \input{figs/intro-map.tikz}
        \caption{
            Map of the environment.
            Agents \protect\agent[Ego] and \protect\agent[Adv] can move synchronously to an adjacent cell (or stay put) at each step.
            Locations \protect\location[EgoHome] and \protect\location[AdvHome] are called power bases.
        }
        \label{fig:ex-capture-idea:map}
    \end{subfigure}%
    \hspace*{0.03\linewidth}%
    \begin{subfigure}[t]{.56\linewidth}
        \centering
        \input{figs/intro-rm-ego.tikz}%
        \caption{
            Reward machine for agent \protect\agent[Ego]
            with sparse reward (zero when not specified on a transition).
            We suppose that \protect\signal[EgoAtHome], \protect\signal[AdvAtHome] and \protect\signal[EgoMeetAdv] cannot all be true at the same time,
            making this reward machine deterministic. $\top$ represents tautology.
        }
        \label{fig:ex-capture-idea:reward-machine}
    \end{subfigure}%
    \caption{
        The simplest variant of the PAC-MAN game is a symmetric zero-sum game where two agents try to capture each other.
        The agent that visited its own power base the most recently can capture the other agent.
        The low-level stochastic game plays out in a grid world (\subref{fig:ex-capture-idea:map}) and
        the high-level reward is captured by (\subref{fig:ex-capture-idea:reward-machine}).
    }
    \label{fig:ex-capture-idea}
    
\end{figure}
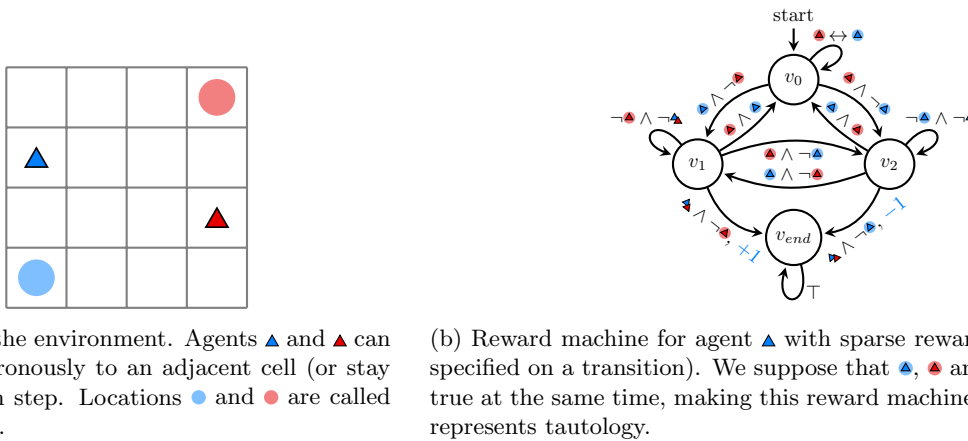

We introduce a variant of the PAC-MAN game as a motivational example,
where two agents try to capture each other,
and the task completion is defined based on complex conditions, as illustrated in \Cref{fig:ex-capture-idea}. 
%
Two agents \agent[Ego] and \agent[Adv] evolve in a grid-world by taking synchronous steps towards adjacent cells. 
We also introduce two power bases \location[EgoHome] and \location[AdvHome] at fixed locations. 
We track whether the two agents encounter logical propositions such as \signal[EgoAtHome], \signal[AdvAtHome], and \signal[EgoMeetAdv] or not to evaluate the lower-level dynamics and specify the high-level reward.
For instance, the proposition \signal[EgoAtHome] is true when agent \agent[Ego] is at power base \location[EgoHome].
Similarly, the proposition \signal[EgoMeetAdv] is true when agents \agent[Ego] and \agent[Adv] meet on the same cell.
In the simplest variant of the game, the goal of each agent is to first reach its own power base and then capture the other agent. Each agent obtains a reward of 1 when its task is completed.
Whenever agent \agent[Ego] arrives at its power base \location[EgoHome], agent \agent[Ego] becomes more powerful than agent \agent[Adv].
However, if agent \agent[Adv] then arrives at its power base \location[AdvHome], agent \agent[Adv] becomes the more powerful agent.
The more powerful agent is capable to capture the other agent.



We address the challenge of learning complex tasks in two-agent general-sum stochastic games with non-Markovian reward functions.
We develop an algorithm called \emph{Q-learning with reward machines for stochastic games (\algoName{})}, where we use reward machines to specify the tasks and expose the structure of reward functions. 
The proposed approach defines the \emph{Q-function at a Nash equilibrium} in \emph{augmented state space} to adapt Q-learning to the setting of stochastic games when the tasks are specified by reward machines. 
The augmented state space integrates the state of the stochastic game and the state of reward machines.
During learning, we formulate a stage game at each time step based on the current Q-functions in augmented state space and use the Lemke-Howson method~\cite{lemke1964equilibrium} to derive a Nash equilibrium. 
Q-functions in augmented state space are then updated according to the Nash equilibrium. 
\algoName{} enables the learning of the best-response strategy at a Nash equilibrium for each agent. 
Q-functions learned in \algoName{} converge to the Q-functions at a Nash equilibrium if the stage game at each time step during learning has a global optimum point or a saddle point, and the agents update Q-functions based on the best-response strategy at this point. 
We test \algoName{} in three case studies and compare \algoName{} with common baselines such as Nash Q learning~\cite{hu2003nash}, Nash Q-learning in augmented state space, multi-agent deep deterministic policy gradient (MADDPG)~\cite{lowe2017multi}, and MADDPG in augmented state space.
The results show that \algoName{} learns the best-response strategies at a Nash equilibrium effectively.

%% file: figs/intro-map.tikz
\begin{tikzpicture}[
    scale=1.6,
    thick,
    every node/.append style={transform shape},
]

\begin{scope}[
    xscale=.5,
    yscale=.5, 
    every node/.append style={scale=2}, 
]

\draw[step=1,color=gray,shift={(-.5,-.5)}] (0,0) grid +(4,4);

\node at (0,0) {\Large\location[EgoHome]};
\node at (3,3) {\Large\location[AdvHome]};
\node at (0,2) {\agent[Ego]};
\node at (3,1) {\agent[Adv]};

\end{scope}

\end{tikzpicture}

%% file: figs/intro-rm-ego.tikz
\begin{tikzpicture}[
    scale=.75,
    thick,
    every node/.append style={transform shape},
]
\node[state,initial above] (uBal) 
    at ( 0,-0)
    {$\mealyCommonState_{0}$};
\node[state] (uEgo)
    at (-1.7,-1.5)
    {$\mealyCommonState_{1}$};
\node[state] (uAdv)
    at (+1.7,-1.5)
    {$\mealyCommonState_{2}$};
\node[state] (uEnd)
    at ( 0,-2.8)
    {$\mealyCommonState_{end}$};

\path[->,sloped]

(uBal) 
edge[out=50,in=20,loop] node[sloped=false,above]
    {$\signal[AdvAtHome]\lequiv\signal[EgoAtHome]$}
    (uBal)
edge[bend right] node[]
    {$\signal[EgoAtHome]\land\lnot\signal[AdvAtHome]$}
    (uEgo)
edge[bend left] node[]
    {$\signal[AdvAtHome]\land\lnot\signal[EgoAtHome]$}
    (uAdv)

(uEgo) 
edge[out=130,in=160,loop] node[sloped=false,above,pos=0.4,xshift=-.5em]
    {$\lnot\signal[AdvAtHome]\land\lnot\signal[EgoMeetAdv]$}
    (uEgo)
edge[bend right=10] node[]
    {$\signal[AdvAtHome]\land\signal[EgoAtHome]$}
    (uBal)
edge[bend left=20] node[swap]
    {$\signal[AdvAtHome]\land\lnot\signal[EgoAtHome]$}
    (uAdv)
edge[bend right] node[swap]
    {$\signal[EgoMeetAdv]\land\lnot\signal[AdvAtHome]$,
    \textcolor{colorPlayerEgo}{$+1$}}
    (uEnd)

(uAdv) 
edge[out=50,in=20,loop] node[sloped=false,above,pos=0.4,xshift=+.5em]
    {$\lnot\signal[EgoAtHome]\land\lnot\signal[EgoMeetAdv]$}
    (uAdv)
edge[bend left=10] node[]
    {$\signal[EgoAtHome]\land\signal[AdvAtHome]$}
    (uBal)
edge[bend left=20] node[]
    {$\signal[EgoAtHome]\land\lnot\signal[AdvAtHome]$} (uEgo)
edge[bend left] node[swap]
    {$\signal[EgoMeetAdv]\land\lnot\signal[EgoAtHome]$,
    \textcolor{colorPlayerEgo}{$-1$}}
    (uEnd)

(uEnd) 
edge[out=-75,in=-105,loop] node[sloped=false,xshift=+1em, yshift=1em]
    {$\top$}
    (uEnd)

;

\end{tikzpicture}

%% file: related_work.tex
\section{Related Work}

\noindent\textbf{Multi-agent reinforcement learning with stochastic games:} Our work is closely related to multi-agent reinforcement learning in stochastic games. Many recent works of multi-agent reinforcement learning (MARL) focus on a cooperative setting ~\cite{zhang2018fully,palmer2017lenient,foerster2017stabilising,omidshafiei2017deep,gupta2017cooperative,foerster2018counterfactual,sunehag2017value,rashid2018qmix}, where all the agents share a common reward function. \cite{foerster2017stabilising} addresses the nonstationarity issue introduced by independent Q-learning in multi-agent reinforcement learning by using the importance sampling technique since the multiple agents usually learn concurrently. With the multi-agent actor-critic method as a backbone, \cite{foerster2018counterfactual} proposes counterfactual multi-agent (COMA) policy gradients that use a centralized critic to estimate the Q-function and decentralized actors to optimize the agents’ policies. Similar to both of these approaches that describe the multi-agent task in the stochastic game, but we tackle a non-cooperative multi-agent reinforcement learning problem in a stochastic game compared to ~\cite{Zhang2019Non-Cooperative,lin2019multi,zhang2021mfvfd,mguni21a}. As opposed to~\cite{Zhang2019Non-Cooperative,lin2019multi}, which extends inverse reinforcement learning (IRL) to the non-cooperative stochastic game to let the agent learn the reward function from the observation of other agent/expert's behavior, we assume the reward function is given to all agents that participate in the stochastic game. However, our work shares the same objective as the works mentioned above, as each agent learns a policy at a Nash equilibrium in the stochastic game. Furthermore, the existing works rely on the Markovian property of the reward function. The definition of the stochastic game in this paper differs from the aforementioned works as we specify the reward function as non-Markovian, which is more applicable to many real-world applications.

\noindent\textbf{Reinforcement learning with formal methods:} This paper is also closely related to the work of using the formal methods in reinforcement learning (RL) with a non-Markovian reward function for single agent ~\cite{hasanbeig2021deepsynth,xu2020joint,toro2019learning,rens2020online,xu2021active,rens2020learning,neider2021advice,de2020restraining,de2020temporal}, and multi-agent \cite{Muniraj2018,leon2020}. In the single-agent RL, \cite{hasanbeig2021deepsynth} proposes DeepSynth which is a new algorithm that synthesizes deterministic finite automation to infer the unknown non-Markovian rewards of achieving a sequence of high-level objectives. However, DeepSynth is based on the general Markov decision process, which makes it incapable of solving MARL problems that are modeled as a stochastic game. In a similar case for multi-agent reinforcement learning, \cite{Muniraj2018} tackles the MARL problem in a stochastic game with non-Markovian reward expressed in temporal logic. \cite{leon2020} incorporate the non-Markovian reward to MARL in an unknown environment, with the complex task specification also described in temporal logic. In this work, the reward function for all the agents in a stochastic game is encoded by the reward machine, which provides an automata-based representation that enables an agent to decompose an RL problem into structured subproblems that can be efficiently learned via off-policy learning\cite{icarte2018}. 

%% file: preliminaries.tex
\section{Preliminaries}
\label{sec:preliminaries}
In this section, we introduce the necessary background on stochastic games, reward machines, and how to connect stochastic games and reward machines.

\subsection{Stochastic Games}

\begin{definition}
\label{def:mdp}
A \emph{two-player general-sum stochastic game} (SG) is a tuple
$\Autom[sg] = (\sgStates, \sgInit, \sgActions_{\xAgent[ego]}, \sgActions_{\xAgent[adv]}, \sgProb, \sgRewardFunction_{\xAgent[ego]}, \sgRewardFunction_{\xAgent[adv]}, \gamma)$, where $\sgStates$ is the finite state space (consisting of states $\sgCommonState=[\sgCommonState_{\xAgent[ego]}; \sgCommonState_{\xAgent[adv]}]$, $\sgCommonState_{\xAgent[ego]}$ and $\sgCommonState_{\xAgent[adv]}$ are the substates of the ego agent and the adversarial agent, respectively), $\sgInit\in\sgStates$ is the initial state,
$\sgActions_{\xAgent[ego]}$ and $\sgActions_{\xAgent[adv]}$ are the finite set of actions for the ego agent and the adversarial agent, respectively,
and 
$\sgProb \colon \sgStates\times \sgActions_{\xAgent[ego]} \times\sgActions_{\xAgent[adv]}\times \sgStates \rightarrow [0,1]$ is the probabilistic transition function.
Reward functions $\sgRewardFunction_{\xAgent[ego]}: (\sgStates \times \sgActions_{\xAgent[ego]}\times\sgActions_{\xAgent[adv]})^+ \times \sgStates \rightarrow \reals$ and $\sgRewardFunction_{\xAgent[adv]}: (\sgStates \times \sgActions_{\xAgent[ego]}\times\sgActions_{\xAgent[adv]})^+ \times \sgStates \rightarrow \reals$ specify the non-Markovian rewards to the ego and adversarial agents, respectively. $\gamma \in (0,1]$ is the discount factor.
\end{definition}

We use the subscripts $\xAgent[ego]$ and $\xAgent[adv]$ to represent the ego agent and the adversarial agent, respectively. Our definition of the SG differs from the ``usual'' definition used in reinforcement learning (e.g.,~\cite{sutton2018reinforcement}) in that the reward functions $\sgRewardFunction_{\xAgent[ego]}$ and $\sgRewardFunction_{\xAgent[adv]}$ are defined over the whole history (i.e., the rewards are \emph{non-Markovian}). A \emph{trajectory} is a sequence of states and actions $s_{\xTime{0}} \sgCommonAction_{\xAgent[ego],\xTime{0}} \sgCommonAction_{\xAgent[adv],\xTime{0}} s_{\xTime{1}} \ldots s_{\xTime[end]} \sgCommonAction_{\xAgent[ego],\xTime[end]} \sgCommonAction_{\xAgent[adv],\xTime[end]} s_{\xTime[end]+1}$,
with $s_{\xTime{0}} = \sgInit$.
Its corresponding \emph{reward sequence} for agent $\xAgent[1]$ ($\xAgent[1] \in \{\xAgent[ego],\xAgent[adv]\}$) is $\sgRewards_{\xAgent[1],1}\ldots\sgRewards_{\xAgent[1],\xTime[end]}$,
where $\sgRewards_{\xAgent[1],\xTime[]} = R_{\xAgent[1]}(\trajectory{\xTime[]})$, for each $\xTime[] \leq \xTime[end]$. 
For both the ego agent and the adversarial agent, they observe the same trajectory \trajectory{\xTime[end]}. The ego agent achieves a discounted cumulative reward $\sum_{\xTime[]=0}^{\xTime[end]} \sgDiscount^{\xTime[]} r_{\xAgent[ego],\xTime[]}$ for the trajectory \trajectory{\xTime[end]}. Similarly, the adversarial agent achieves a discounted cumulative reward $\sum_{\xTime[]=0}^{\xTime[end]} \sgDiscount^{\xTime[]} r_{\xAgent[adv],\xTime[]}$  for the trajectory \trajectory{\xTime[end]}.

In an SG, we consider each agent to be equipped with a Markovian strategy $\pi_{\xAgent[1]}: S \times A_{\xAgent[1]} \rightarrow [0, 1], \xAgent[1] \in \{\xAgent[ego],\xAgent[adv]\}$, mapping the state to the probability of selecting each possible action. 
The objective of the ego agent and the adversarial agent is to maximize the expected discounted cumulative reward $\Tilde{v}_{\xAgent[ego]}$ and $\Tilde{v}_{\xAgent[adv]}$, respectively.
\begin{equation}
    \Tilde{v}_{\xAgent[1]}(s, \pi_{\xAgent[ego]}, \pi_{\xAgent[adv]}) = \sum_{\xTime[]=0}^{\infty} \gamma^{\xTime[]} \mathbb{E}(r_{\xAgent[1],\xTime[]}|\pi_{\xAgent[ego]}, \pi_{\xAgent[adv]}, s_{\xTime{0}} = s)
\end{equation}

As a solution concept of an SG, a Nash equilibrium \cite{nash1951non} is a collection of strategies for each of the agents such that each agent cannot improve its own reward by changing its own strategy, or we say that, each agent's strategy is a \textit{best-response} to the other agents’ strategies. 
\begin{definition}
For an SG $\Autom[sg] = (\sgStates, \sgInit, \sgActions_{\xAgent[ego]}, \sgActions_{\xAgent[adv]}, R_{\xAgent[ego]}, R_{\xAgent[adv]}, \sgProb, \gamma)$, the strategies of the ego agent and the adversarial agent $\pi^{\ast}_{\xAgent[ego]}$ and $\pi^{\ast}_{\xAgent[adv]}$ are at a Nash equilibrium of the SG if
\[
\begin{split}
    \Tilde{v}_{\xAgent[ego]}(s, \pi^{\ast}_{\xAgent[ego]}, \pi^{\ast}_{\xAgent[adv]})&\ge \Tilde{v}_{\xAgent[ego]}(s, \pi_{\xAgent[ego]}, \pi^{\ast}_{\xAgent[adv]}) \\
    \Tilde{v}_{\xAgent[adv]}(s, \pi^{\ast}_{\xAgent[ego]}, \pi^{\ast}_{\xAgent[adv]})&\ge \Tilde{v}_{\xAgent[adv]}(s, \pi^{\ast}_{\xAgent[ego]}, \pi_{\xAgent[adv]})
\end{split}
\]
hold for any $\pi_{\xAgent[ego]}$ and $\pi_{\xAgent[adv]}$.
\end{definition}
A Nash equilibrium illustrates the stable point where each agent reacts optimally to the behavior of other agents. In a Nash equilibrium, given other agents' strategies, the learning agent is not able to get a higher reward by unilaterally deviating from its current strategy. We refer to $\pi^{\ast}_{\xAgent[1]}$ as the strategy of agent $\xAgent[1]$ that constitutes a Nash equilibrium. The goal of each learning agent is to find a strategy that maximizes its discounted cumulative reward considering the other agent's action once it reaches the game's Nash equilibrium. 

\subsection{Reward Machines}
In this paper, we use reward machines to express the task specification. Reward machines \cite{DBLP:conf/icml/IcarteKVM18,LTLAndBeyond} encode a (non-Markovian) reward in a type of finite-state machine. 
\footnote{
	The reward machines we are utilizing are the so-called \emph{simple reward machines} in the parlance of~\cite{DBLP:conf/icml/IcarteKVM18},
	where every output symbol is a real number.
}
Technically, a reward machine is a special instance of a Mealy machine~\cite{DBLP:books/daglib/0025557},
the one that takes subsets of propositional variables as its input and outputs real numbers as reward values.

\begin{definition}
\label{def:rewardMealyMachines}
A \emph{reward machine} (RM)
$\Autom[rm] = (\mealyStates, \mealyInit, \mealyInputAlphabet, \mealyOutputAlphabet, \mealyTransition, \mealyOutput)$ consists of
a finite, nonempty set $\mealyStates$ of RM states, 
an initial RM state $\mealyInit \in \mealyStates$, 
an input alphabet $\mealyInputAlphabet$ where $\rmLabels$ is a finite set of propositional variables,
an output alphabet $\mealyOutputAlphabet$, 
a (deterministic) transition function $\mealyTransition \colon \mealyStates \times \mealyInputAlphabet \to \mealyStates$, 
and an output function $\mealyOutput \colon \mealyStates \times \mealyInputAlphabet \to \mealyOutputAlphabet$.
\end{definition}

By using reward machines, we decompose a complex task into several stages. For example, considering the reward machine of the ego agent (\agent[Ego]) in the motivational example (\Cref{fig:ex-capture-idea:reward-machine}), the initial RM state is $v_{\xTime{0}}$. At $v_{\xTime{0}}$, if the proposition \signal[EgoAtHome] becomes true and \signal[AdvAtHome] is not reached, the machine transitions to $v_{\xTime{1}}$ and outputs a reward of 0, or if the proposition \signal[AdvAtHome] becomes true and \signal[EgoAtHome] is not reached, the machine transitions to $v_2$ and outputs a reward of 0. At $v_{\xTime{1}}$, after \signal[EgoMeetAdv] is reached and the adversarial agent is not at its power base ($\lnot \signal[AdvAtHome]$), the ego agent completes its task, and the machine transitions to $v_{end}$ and outputs a reward of 1. Similarly, at $v_2$, after \signal[EgoMeetAdv] is reached and $\lnot$\signal[EgoAtHome] is true, the machine transitions to $v_{end}$ and outputs a reward of -1. 
Additionally, as any agent can be the more powerful of the two (i.e. with the power to capture the other agent) once it arrives at its own power base, the machine can transition between $v_{\xTime{1}}$ and $v_2$. 
We also have self-loops if the propositions do not lead to a change in the stage of the task.

\subsection{Connecting Stochastic Games and Reward Machines}
To build a bridge between the reward machine and the SG, a labeling function $L: S \times A_{\xAgent[ego]} \times A_{\xAgent[adv]} \times S \rightarrow 2^{\rmLabels}$ is required to map the states in an SG to the \emph{high-level events}. High-level events represent expert knowledge of what is relevant for successfully executing a task and are assumed to be available to both the ego and adversarial agents.
At time step $\xTime[]$, we have $l_{\xTime[]} = L(s_{\xTime[]}, \sgCommonAction_{\xAgent[ego],\xTime[]}, \sgCommonAction_{\xAgent[adv],\xTime[]}, s_{\xTime{\xTime[]+1}})$ to determine the set of relevant high-level events that the agents detect in the environment. $l_{\xTime[]}$ is a set consisting of the propositions in $\rmLabels$ that are true given $(s_{\xTime[]},\sgCommonAction_{\xAgent[ego],\xTime[]}, \sgCommonAction_{\xAgent[adv],\xTime[]}, s_{\xTime{\xTime[]+1}})$. 
We assume that the agents share the same set of propositional variables $\rmLabels$, and the agents observe the same sequences of high-level events for the same trajectory. Specifically, for the trajectory $\trajectory{\xTime[end]}$, its corresponding \emph{event sequence} is $l_{\xTime{0}} l_{\xTime{1}} \ldots l_{\xTime[end]}$. 
The reward machine $\Autom[rm]$ receives $l_{\xTime[]}$ as an input and outputs a reward $r_{\xTime[]} = \mealyOutput(v_{\xTime[]}, l_{\xTime[]})$. 
To connect the input and output of $\Autom[rm]$, we write $\Autom[rm](l_{\xTime{0}}l_{\xTime{1}} \cdots l_{\xTime[end]}) = r_{\xTime{0}}r_{\xTime{1}}\cdots r_{\xTime[end]} $.  
We call $(l_{\xTime{0}} l_{\xTime{1}} \ldots l_{\xTime[end]}, r_{\xTime{0}}r_{\xTime{1}}\cdots r_{\xTime[end]})$  a trace and we consider finite traces (with possibly unbounded length) in this paper as a reward machine maps any sequence of events to a sequence of rewards. 
Given the detected high-level event $l_{\xTime[]}$ at time step $\xTime[]$, the RM state transitions to $v_{\xTime{\xTime[]+1}} = \mealyTransition(v_{\xTime[]},l_{\xTime[]})$.

\begin{definition}
A reward machine $\Autom[rm]_{\xAgent[1]} = (\mealyStates_{\xAgent[1]}, v_{I,\xAgent[1]}, \mealyInputAlphabet, \mealyOutputAlphabet_{\xAgent[1]}, \mealyTransition_{\xAgent[1]}, \mealyOutput_{\xAgent[1]})$ $(\xAgent[1] \in \{\xAgent[ego],\xAgent[adv]\})$ encodes the non-Markovian reward function $R_{\xAgent[1]}$ of the agent $\xAgent[1]$ in the SG $\Autom[sg] = (\sgStates, \sgInit, \sgActions_{\xAgent[ego]}, \sgActions_{\xAgent[adv]}, R_{\xAgent[ego]}, R_{\xAgent[adv]}, \sgProb, \gamma)$, if for every trajectory $\trajectory{\xTime[end]}$ and the corresponding event sequence $l_{\xTime{0}} l_{\xTime{1}} \cdots l_{\xTime[end]}$,
the reward sequence $r_{\xAgent[1],\xTime{0}} \cdots r_{\xAgent[1],\xTime[end]}$ of the agent $\xAgent[1]$ equals $\Autom[rm]_{\xAgent[1]}(l_{\xTime{0}} l_{\xTime{1}} \cdots l_{\xTime[end]})$.
\end{definition}

We use reward machines $\Autom[rm]_{\xAgent[ego]} \text{ and } \Autom[rm]_{\xAgent[adv]}$ to encode the non-Markovian reward functions $R_{\xAgent[ego]} \text{ and } R_{\xAgent[adv]}$ in the SG $\Autom[sg]$ for the ego agent and the adversarial agent, respectively.




%% file: problem.tex
\section{Problem formulation}

We focus on a two-agent general-sum stochastic game (SG) with non-Markovian reward functions, $\Autom[sg] = (\sgStates, \sgInit,  \sgActions_{\xAgent[ego]},\\ \sgActions_{\xAgent[adv]}, R_{\xAgent[ego]}, R_{\xAgent[adv]}, \sgProb, \gamma)$. Each agent is given a task to complete, and the task completion depends on the other agent's behavior. 
We use a separate reward machine $\Autom[rm]_{\xAgent[1]}=(\mealyStates_{\xAgent[1]}, v_{I,\xAgent[1]}, \mealyInputAlphabet, \mealyOutputAlphabet, \mealyTransition_{\xAgent[1]}, \mealyOutput_{\xAgent[1]})$ for agent $\xAgent[1]$ ($\xAgent[1] \in \{\xAgent[ego],\xAgent[adv]\}$) to specify a task that the agent intends to achieve. Moreover, $\Autom[rm]_{\xAgent[1]}$ encodes the non-Markovian reward function $R_{\xAgent[1]}$ in $\Autom[sg]$. An agent obtains a high discounted cumulative reward if its task is completed. Therefore, seeking to accomplish the task is consistent with maximizing the discounted cumulative reward for the learning agent.

In the two-player general-sum stochastic game, the agents operate in an adversarial environment, e.g., in the motivational example, the task of the ego agent is accomplished if the ego agent captures the adversarial agent. Similarly, the task of the adversarial agent is accomplished if the adversarial agent captures the ego agent.

In this paper, we aim to find the best-response strategy for each agent in the two-player general-sum stochastic game with reward functions encoded by reward machines. Agents try to maximize their own discounted cumulative reward.
We have the following problem formulation.

\begin{problem}
Given an SG $\Autom[sg] = (\sgStates, \sgInit, \sgActions_{\xAgent[ego]}, \sgActions_{\xAgent[adv]}, R_{\xAgent[ego]}, R_{\xAgent[adv]}, \sgProb, \gamma)$, where the non-Markovian reward functions $R_{\xAgent[ego]}$ and $R_{\xAgent[adv]}$ are encoded by reward machines $\Autom[rm]_{\xAgent[ego]} = (\mealyStates_{\xAgent[ego]}, v_{I,\xAgent[ego]}, 2^{\rmLabels}, \mealyOutputAlphabet_{\xAgent[ego]}, \mealyTransition_{\xAgent[ego]}, \mealyOutput_{\xAgent[ego]})$ and $\Autom[rm]_{\xAgent[adv]} = (\mealyStates_{\xAgent[adv]}, v_{I,\xAgent[adv]}, 2^{\rmLabels}, \mealyOutputAlphabet_{\xAgent[adv]}, \mealyTransition_{\xAgent[adv]}, \mealyOutput_{\xAgent[adv]})$, respectively, learn the strategies of the ego agent and the adversarial agent $\pi^{\ast}_{\xAgent[ego]}$ and $\pi^{\ast}_{\xAgent[adv]}$ at a Nash equilibrium of the stochastic game.
\end{problem}
We assume that the state, RM state, selected action, and earned reward of both agents are observable to each agent. The set of propositional variables $\rmLabels$ is the same in $\Autom[rm]_{\xAgent[ego]}$ and $\Autom[rm]_{\xAgent[adv]}$.
The high-level events received in the reward machines are a function of both agents' states and actions. Therefore, the learning processes of the two agents are coupled.

%% file: solution.tex
\section{Q-learning with Reward Machine for Stochastic Game (\algoName)}
\label{sec_baseline}




In this section, we introduce the proposed {\algoName} algorithm for stochastic games with reward functions encoded by reward machines. We first define a stochastic game with reward machines and formulate it as a product stochastic game to obtain Markovian rewards. Then, we use {\algoName} to learn a strategy for each agent that constitutes a Nash equilibrium.

\begin{definition}
Given a SG $\Autom[sg] = (\sgStates, \sgInit, \sgActions_{\xAgent[ego]}, \sgActions_{\xAgent[adv]}, R_{\xAgent[ego]}, R_{\xAgent[adv]}, \sgProb, \gamma)$, where the reward functions $R_{\xAgent[ego]}$ and $R_{\xAgent[adv]}$ are encoded by reward machines $\Autom[rm]_{\xAgent[ego]} = (\mealyStates_{\xAgent[ego]}, v_{I,\xAgent[ego]}, 2^{\rmLabels}, \mealyOutputAlphabet_{\xAgent[ego]}, \mealyTransition_{\xAgent[ego]}, \mealyOutput_{\xAgent[ego]})$ and $\Autom[rm]_{\xAgent[adv]} = (\mealyStates_{\xAgent[adv]}, v_{I,\xAgent[adv]}, 2^{\rmLabels}, \mealyOutputAlphabet_{\xAgent[adv]}, \mealyTransition_{\xAgent[adv]}, \mealyOutput_{\xAgent[adv]})$, respectively,
we define a stochastic game with reward machines (SGRM) as a product stochastic game $\Autom[sgrm] = (\sgStates', \sgInit', \sgActions_{\xAgent[ego]}, \sgActions_{\xAgent[adv]}, \sgRewardFunction_{\xAgent[ego]}', \sgRewardFunction_{\xAgent[adv]}', \sgProb', \gamma)$, where
\begin{itemize}
    \item $\sgStates'=\sgStates \times V_{\xAgent[ego]} \times V_{\xAgent[adv]}$, 
    \item $\sgInit' = \sgInit \times v_{I,\xAgent[ego]} \times v_{I,\xAgent[adv]}$ ,
    \item $\sgProb'(s, v_{\xAgent[ego]}, v_{\xAgent[adv]}, a_{\xAgent[ego]}, a_{\xAgent[adv]}, s', v'_{\xAgent[ego]}, v'_{\xAgent[adv]}) = 
        \begin{cases}
            \sgProb(s, \sgCommonAction_{\xAgent[ego]}, \sgCommonAction_{\xAgent[adv]}, s') & \text{if } v'_{\xAgent[ego]}=\mealyTransition_{\xAgent[ego]}(v_{\xAgent[ego]}, L(s, \sgCommonAction_{\xAgent[ego]}, \sgCommonAction_{\xAgent[adv]}, s')) \\ & \text{and } v'_{\xAgent[adv]}=\mealyTransition_{\xAgent[adv]}(v_{\xAgent[adv]}, L(s,\sgCommonAction_{\xAgent[ego]},\sgCommonAction_{\xAgent[adv]},s')),\\
            0 & \text{otherwise,} \\
        \end{cases}$ \\
    \item $\sgRewardFunction_{\xAgent[ego]}' = \mealyOutput_{\xAgent[ego]}(v_{\xAgent[ego]}, L(s, \sgCommonAction_{\xAgent[ego]}, \sgCommonAction_{\xAgent[adv]}, s'))$,
    \item $\sgRewardFunction_{\xAgent[adv]}' = \mealyOutput_{\xAgent[adv]}(v_{\xAgent[adv]}, L(s,\sgCommonAction_{\xAgent[ego]},\sgCommonAction_{\xAgent[adv]},s'))$.
\end{itemize}
where $L$ is the labeling function $L:S \times A_{\xAgent[ego]} \times A_{\xAgent[adv]} \times S \rightarrow 2^{\rmLabels}$.
\end{definition}

\begin{lemma}
A stochastic game with reward machines (SGRM) has Markovian reward functions to express the non-Markovian reward functions in the stochastic game encoded by reward machines. 
\end{lemma}

The agents consider $\sgStates'$ to select actions. At each time step, the agents select actions simultaneously and execute the actions $\sgCommonAction_{\xAgent[ego]}$ and $\sgCommonAction_{\xAgent[adv]}$ respectively. 
Then the state $s$ moves to $s'$. The labeling function $L$ returns the high-level events given current state and actions, and next state. Given the events, reward machines deliver the transition of RM states, $(v_{\xAgent[ego]}, v_{\xAgent[adv]})$ moving to $(v'_{\xAgent[ego]}, v'_{\xAgent[adv]})$, where $v'_{\xAgent[ego]}=\mealyTransition_{\xAgent[ego]}(v_{\xAgent[ego]}, L(s, \sgCommonAction_{\xAgent[ego]}, \sgCommonAction_{\xAgent[adv]}, s')),$ $ v'_{\xAgent[adv]}=\mealyTransition_{\xAgent[adv]}(v_{\xAgent[adv]}, L(s, \sgCommonAction_{\xAgent[ego]}, \sgCommonAction_{\xAgent[adv]},s'))$. The transition ($(s, v_{\xAgent[ego]}, v_{\xAgent[adv]})$ to $(s,v'_{\xAgent[ego]},v'_{\xAgent[adv]})$) leads to a reward of $\mealyOutput_{\xAgent[ego]}(v_{\xAgent[ego]}, L(s, \sgCommonAction_{\xAgent[ego]}, \sgCommonAction_{\xAgent[adv]}, s'))$ for the ego agent and $\mealyOutput_{\xAgent[adv]}(v_{\xAgent[adv]}, L(s, \sgCommonAction_{\xAgent[ego]}, \sgCommonAction_{\xAgent[adv]},s'))$ for the adversarial agent.

To utilize Q-learning and considering a Nash equilibrium as a solution concept, we consider the optimal Q-function for agent $\xAgent[1]$ ($\xAgent[1] \in \{\xAgent[ego],\xAgent[adv]\}$) in an SGRM is the \emph{Q-function at a Nash equilibrium} inspired by \cite{hu2003nash}, which is the expected discounted cumulative reward obtained by agent $\xAgent[1]$ when both agents follow a joint Nash equilibrium strategy from the next period on. Therefore, Q-function at a Nash equilibrium depends on both agent's actions. Moreover, we define the Q-function at a Nash equilibrium in augmented state space $S\times V_{\xAgent[ego]}\times V_{\xAgent[adv]}$. 
Let $q^{\ast}_{\xAgent[1]}(s, v_{\xAgent[ego]}, v_{\xAgent[adv]}, \sgCommonAction_{\xAgent[ego]}, \sgCommonAction_{\xAgent[adv]})$ denote the Q-function at a Nash equilibrium for agent $\xAgent[1]$.
Mathematically,
\begin{equation}
    q^{\ast}_{\xAgent[1]}(s, v_{\xAgent[ego]}, v_{\xAgent[adv]}, \sgCommonAction_{\xAgent[ego]}, \sgCommonAction_{\xAgent[adv]})
    =
    r_{\xAgent[1]}(s, v_{\xAgent[ego]}, v_{\xAgent[adv]}, \sgCommonAction_{\xAgent[ego]}, \sgCommonAction_{\xAgent[adv]})
    + \gamma \sum_{\substack{s' \in S \\ v'_{\xAgent[ego]} \in V_{\xAgent[ego]} \\ v'_{\xAgent[adv]} \in V_{\xAgent[adv]}}} p(s',v'_{\xAgent[ego]},v'_{\xAgent[adv]}|s,v_{\xAgent[ego]},v_{\xAgent[adv]}, \sgCommonAction_{\xAgent[ego]}, \sgCommonAction_{\xAgent[adv]}) \Tilde{v}_{\xAgent[1]}(s', v'_{\xAgent[ego]},v'_{\xAgent[adv]},\pi_{\xAgent[ego]}^\ast, \pi_{\xAgent[adv]}^\ast)
\end{equation}
where $\Tilde{v}_{\xAgent[1]}(s', v'_{\xAgent[ego]},v'_{\xAgent[adv]},\pi_{\xAgent[ego]}^\ast, \pi_{\xAgent[adv]}^\ast)$ is the expected discounted cumulative reward starting from the augmented state $(s',v'
_{\xAgent[ego]},v'_{\xAgent[adv]})$ over infinite periods when agents follow the best-response strategies $\pi_{\xAgent[ego]}^{\ast}$ and $\pi_{\xAgent[adv]}^{\ast}$.
When multiple equilibria are derived, different Nash strategy profile may lead to different Q-function at a Nash equilibrium.

The proposed algorithm {\algoName} tries to learn $q^{\ast}_{\xAgent[1]}$ ($\xAgent[1] \in \{\xAgent[ego],\xAgent[adv]\}$) for a Nash equilibrium strategy. At a Nash equilibrium, the strategy of one agent is optimal when considering the other agent's behavior. 
Each agent maintains two Q-functions --- learning the Q-functions of both itself and the other agent. $q_{\xAgent[1]\xAgent[2]}$ represents the Q-function for agent $\xAgent[1]$ and learned by agent $\xAgent[2]$. For instance, the ego agent is equipped with $q_{\xAgent[ego]\xAgent[ego]}$ and $q_{\xAgent[adv]\xAgent[ego]}$. Similarly, $\pi_{\xAgent[1]\xAgent[2]}$ is the strategy of agent $\xAgent[1]$ learned by agent $\xAgent[2]$.
The agents learn the Q-functions through exploration and by observing state and RM state transitions, actions, and rewards.
At each time step during learning, we can formulate a \emph{stage game} for agent $\xAgent[1]$ given the Q-functions $q_{\xAgent[ego]\xAgent[1]}(s,v_{\xAgent[ego]},v_{\xAgent[adv]})$ and $q_{\xAgent[adv]\xAgent[1]}(s,v_{\xAgent[ego]},v_{\xAgent[adv]})$ estimated by agent $\xAgent[1]$. 
\begin{definition}
A two-player stage game is defined as $(r_{\xAgent[ego]},r_{\xAgent[adv]})$, where $r_{\xAgent[1]}$ $(\xAgent[1] \in \{\xAgent[ego],\xAgent[adv]\})$ is agent $\xAgent[1]$’s reward function $R_{\xAgent[1]}$ over the space of joint actions, $r_{\xAgent[1]} = \{R_{\xAgent[1]}(\sgCommonAction_{\xAgent[ego]}, \sgCommonAction_{\xAgent[adv]})|\sgCommonAction_{\xAgent[ego]} \in A_{\xAgent[ego]}, \sgCommonAction_{\xAgent[adv]} \in A_{\xAgent[adv]}\}$.
\end{definition}
\noindent Given the stage game ($q_{\xAgent[ego]\xAgent[1]}(s,v_{\xAgent[ego]},v_{\xAgent[adv]}),q_{\xAgent[adv]\xAgent[1]}(s,v_{\xAgent[ego]},v_{\xAgent[adv]})$) during learning, we derive the best-response strategies at a Nash equilibrium and Q-functions are updated based on the expectation that agents would take best-response actions.


\begin{algorithm}[!h] 
	\caption{QRM-SG}
	\label{alg:QRMSG}
	\DontPrintSemicolon
	\SetKwBlock{Begin}{function}{end function}
	{ \textbf{Hyperparameter}: episode length \textit{eplength}, $\gamma$, $\epsilon$ } \;
	{ \textbf{Input:} Reward machines $\Autom[rm]_{\xAgent[ego]}, \Autom[rm]_{\xAgent[adv]}$}\;
	{$\sgCommonState \gets \mathit{InitialState()}; \mealyCommonState_{\xAgent[ego]} \gets v_{I,\xAgent[ego]}; \mealyCommonState_{\xAgent[adv]} \gets v_{I,\xAgent[adv]}$ }\; 
	{$
 q_{\xAgent[ego]\xAgent[ego]}(s, v_{\xAgent[ego]}, v_{\xAgent[adv]}, \sgCommonAction_{\xAgent[ego]}, \sgCommonAction_{\xAgent[adv]}),\allowbreak
 q_{\xAgent[adv]\xAgent[ego]}(s, v_{\xAgent[ego]}, v_{\xAgent[adv]}, \sgCommonAction_{\xAgent[ego]}, \sgCommonAction_{\xAgent[adv]}),\allowbreak
 q_{\xAgent[ego]\xAgent[adv]}(s, v_{\xAgent[ego]}, v_{\xAgent[adv]}, \sgCommonAction_{\xAgent[ego]}, \sgCommonAction_{\xAgent[adv]}),\allowbreak
 q_{\xAgent[adv]\xAgent[adv]}(s, v_{\xAgent[ego]}, v_{\xAgent[adv]}, \sgCommonAction_{\xAgent[ego]}, \sgCommonAction_{\xAgent[adv]})\allowbreak
 \gets \mathit{InitialQfunctions()}$ \label{algLine:InitialQ} }\;
 \For{$episode=1,2,\cdots$}
 {
	\For{$\xTime{0} \leq \xTime[] < \mathit{eplength}$ \label{algLine:taskLoopStart}} 
	{
	\For{$\xAgent[1] \in \{\xAgent[ego],\xAgent[adv]\}$ \label{algLine:actionLoop}}
	{
	{$\overline{\epsilon} =$ GenerateRandomValue() \label{algLine:a1}}\;
	\eIf{ $\overline{\epsilon}< \epsilon$ \label{algLine:a2}}
	    {$\sgCommonAction_{\xAgent[1]} \gets $ ChooseActionRandomly()  \label{algLine:a3}\;}{
	    $\pi_{\xAgent[ego]\xAgent[1]}(\cdot|s,v_{\xAgent[ego]},v_{\xAgent[adv]}), \pi_{\xAgent[adv]\xAgent[1]}(\cdot|s,v_{\xAgent[ego]},v_{\xAgent[adv]}) \gets $ CalculateNashEquilibrium($q_{\xAgent[ego]\xAgent[1]}(s, v_{\xAgent[ego]}, v_{\xAgent[adv]}),\allowbreak q_{\xAgent[adv]\xAgent[1]}(s, v_{\xAgent[ego]}, v_{\xAgent[adv]})$) \label{algLine:a5} \\
	    $\sgCommonAction_{\xAgent[1]} \gets \underset{\sgCommonAction \in A_{\xAgent[1]}}{\mathrm{argmax}}\ \pi_{ii}(\sgCommonAction|s, v_{\xAgent[ego]}, v_{\xAgent[adv]})$   \label{algLine:a6}}
	}
	    { $\sgCommonState' \gets \text{ExecuteAction}(\sgCommonState, \sgCommonAction_{\xAgent[ego]}, \sgCommonAction_{\xAgent[adv]})$ \label{algLine:executeAction}} \;
	    {$l_{\xTime[]} \gets \rmLabelingFunction(\sgCommonState, \sgCommonAction_{\xAgent[ego]}, \sgCommonAction_{\xAgent[adv]}, s')$ \label{algLine:RMlabel}} \;
	    \For{$\xAgent[1] \in \{\xAgent[ego],\xAgent[adv]\}$ \label{algLine:rmLoop}}
	    {
		{  $\mealyCommonState'_{\xAgent[1]} \gets \mealyTransition_{\xAgent[1]}( \mealyCommonState_{\xAgent[1]}, l_{\xTime[]}) $ \label{algLine:RMTransition} } \;
		{$r_{\xAgent[1],\xTime[]} \gets \mealyOutput_{\xAgent[1]}(\mealyCommonState_{\xAgent[1]}, l_{\xTime[]}) $ \label{algLine:Reward}}  \;
		}
		\For{$\xAgent[1] \in \{\xAgent[ego],\xAgent[adv]\}$ \label{algLine:taskLoopStart2}}  
	    {
	    {$\pi_{\xAgent[ego]\xAgent[1]}(\cdot|s', v'_{\xAgent[ego]}, v'_{\xAgent[adv]}), \pi_{\xAgent[adv]\xAgent[1]}(\cdot|s', v'_{\xAgent[ego]}, v'_{\xAgent[adv]}) \gets $ CalculateNashEquilibrium($q_{\xAgent[ego]\xAgent[1]}(s', v'_{\xAgent[ego]}, v'_{\xAgent[adv]}),\allowbreak q_{\xAgent[adv]\xAgent[1]}(s', v'_{\xAgent[ego]}, v'_{\xAgent[adv]})$) \label{algLine:Qpi}} \;
     {$\overline{q}_{\xAgent[ego]\xAgent[1]}(s', v'_{\xAgent[ego]}, v'_{\xAgent[adv]}) \gets \allowbreak \pi_{\xAgent[ego]\xAgent[1]}(\cdot|s', v'_{\xAgent[ego]}, v'_{\xAgent[adv]})\pi_{\xAgent[adv]\xAgent[1]}(\cdot|s', v'_{\xAgent[ego]}, v'_{\xAgent[adv]})q_{\xAgent[ego]\xAgent[1]}(s',v'_{\xAgent[ego]}, v'_{\xAgent[adv]})$\label{algLine:calNashQ1}} \\
     {$\overline{q}_{\xAgent[adv]\xAgent[1]}(s', v'_{\xAgent[ego]}, v'_{\xAgent[adv]}) \gets \allowbreak \pi_{\xAgent[ego]\xAgent[1]}(\cdot|s', v'_{\xAgent[ego]}, v'_{\xAgent[adv]})\pi_{\xAgent[adv]\xAgent[1]}(\cdot|s', v'_{\xAgent[ego]}, v'_{\xAgent[adv]})q_{\xAgent[adv]\xAgent[1]}(s',v'_{\xAgent[ego]}, v'_{\xAgent[adv]})$ \label{algLine:calNashQ2}} \\
     {$q_{\xAgent[ego]\xAgent[1]}(s, v_{\xAgent[ego]}, v_{\xAgent[adv]}, \sgCommonAction_{\xAgent[ego]}, \sgCommonAction_{\xAgent[adv]}) \gets \allowbreak (1-\alpha_{\xTime[]})q_{\xAgent[ego]\xAgent[1]}(s, v_{\xAgent[ego]}, v_{\xAgent[adv]}, \sgCommonAction_{\xAgent[ego]}, \sgCommonAction_{\xAgent[adv]}) + \alpha_{\xTime[]}(r_{\xAgent[ego],\xTime[]} + \gamma \overline{q}_{\xAgent[ego]\xAgent[1]}(s', v'_{\xAgent[ego]}, v'_{\xAgent[adv]}))$ \label{algLine:updateQ1}} \\   
     {$q_{\xAgent[adv]\xAgent[1]}(s, v_{\xAgent[ego]}, v_{\xAgent[adv]}, \sgCommonAction_{\xAgent[ego]}, \sgCommonAction_{\xAgent[adv]}) \gets (1-\alpha_{\xTime[]})q_{\xAgent[adv]\xAgent[1]}(s, v_{\xAgent[ego]}, v_{\xAgent[adv]}, \sgCommonAction_{\xAgent[ego]}, \sgCommonAction_{\xAgent[adv]}) + \alpha_{\xTime[]}(r_{\xAgent[adv],\xTime[]} + \gamma \overline{q}_{\xAgent[adv]\xAgent[1]}(s', v'_{\xAgent[ego]}, v'_{\xAgent[adv]}))$ \label{algLine:updateQ2}} \\
	    }
		{ $\sgCommonState \gets \sgCommonState'; \mealyCommonState_{\xAgent[ego]} \gets \mealyCommonState'_{\xAgent[ego]}; \mealyCommonState_{\xAgent[adv]} \gets \mealyCommonState'_{\xAgent[adv]}$ \label{algLine:taskLoopEnd}} \\
	}
 }
	{ \Return $(q_{\xAgent[ego]\xAgent[ego]}, q_{\xAgent[adv]\xAgent[ego]},q_{\xAgent[ego]\xAgent[adv]},q_{\xAgent[adv]\xAgent[adv]})$}
\end{algorithm}

\Cref{alg:QRMSG} shows the pseudocode for {\algoName}. {\algoName} begins with initializations of Q-functions (\cref{algLine:InitialQ}). Within an episode, each agent interacts with the environment (\crefrange{algLine:actionLoop}{algLine:Reward}), and uses the observations perceived from the environment to update Q-functions (\crefrange{algLine:taskLoopStart2}{algLine:updateQ2}). The game restarts when the number of time steps reaches the threshold or at least one agent completes the task.

We have three loops within one episode in the {\algoName}. The first loop (\crefrange{algLine:actionLoop}{algLine:a6}) demonstrates how the agents select an action, where $\epsilon-\text{greedy}$ policy \cite{sutton2018reinforcement} is adopted to balance the exploration and exploitation. A random floating point number $\overline{\epsilon}$ is uniformly sampled in the range $[0.0, 1.0)$ (\cref{algLine:a1}). When $\overline{\epsilon} < \epsilon$ which has a probability of $\epsilon$ (\cref{algLine:a2}), the learning agent takes a uniformly random action (\cref{algLine:a3}). With probability $1-\epsilon$, the learning agent takes the Nash equilibrium action (\cref{algLine:a5,algLine:a6}). In implementation, we use the Lemke-Howson method \cite{lemke1964equilibrium} to derive a Nash equilibrium. In \cref{algLine:a5}, the Lemke-Howson algorithm takes the learned Q-functions $q_{\xAgent[ego]\xAgent[1]}(s, v_{\xAgent[ego]}, v_{\xAgent[adv]})$ and $q_{\xAgent[adv]\xAgent[1]}(s, v_{\xAgent[ego]}, v_{\xAgent[adv]})$ with respect to the current augmented state $(s, v_{\xAgent[ego]}, v_{\xAgent[adv]})$ as the input, and returns a Nash equilibrium which specifies probabilities of each available action. Then, the agent selects the action with the maximum probability (\cref{algLine:a6}). In \cref{algLine:executeAction}, agents execute the selected actions and the state $s$ transitions to $s'$ according to the function $p$ in the stochastic game $\Autom[sg]$.

The second loop (\crefrange{algLine:rmLoop}{algLine:Reward}) shows the transitions in the corresponding reward machine for each agent. The labeling function first detects the high-level event (\cref{algLine:RMlabel}). Given the current RM state and detected event, for each agent, we track the transition of RM state (\cref{algLine:RMTransition}) and compute the reward that the agent would receive according to its reward machine (\cref{algLine:Reward}). 

The third loop (\crefrange{algLine:taskLoopStart2}{algLine:updateQ2}) is the learning loop. For agent $\xAgent[1]$, a Nash equilibrium at $(s', v'_{\xAgent[ego]}, v'_{\xAgent[adv]})$ is first computed using the Q-functions learned by agent $\xAgent[1]$ (\cref{algLine:Qpi}). Then in \cref{algLine:calNashQ1,algLine:calNashQ2}, we calculate the discounted cumulative reward $\overline{q}_{\xAgent[2]\xAgent[1]}$ of agent $\xAgent[2]$ $(\xAgent[2] \in \{\xAgent[ego],\xAgent[adv]\})$ when the agents follow the Nash equilibrium obtained in \cref{algLine:Qpi} at $(s', v'_{\xAgent[ego]}, v'_{\xAgent[adv]})$. $\overline{q}_{\xAgent[2]\xAgent[1]}$ uses the corresponding Q-functions $q_{\xAgent[2]\xAgent[1]}$ to derive the discounted cumulative reward at $(s', v'_{\xAgent[ego]}, v'_{\xAgent[adv]})$.
\begin{equation}
    \overline{q}_{\xAgent[2]\xAgent[1]}(s', v'_{\xAgent[ego]}, v'_{\xAgent[adv]}) = \pi_{\xAgent[ego]\xAgent[1]}(s', v'_{\xAgent[ego]}, v'_{\xAgent[adv]})\pi_{\xAgent[adv]\xAgent[1]}(s', v'_{\xAgent[ego]}, v'_{\xAgent[adv]})q_{\xAgent[2]\xAgent[1]}(s',v'_{\xAgent[ego]}, v'_{\xAgent[adv]})
\end{equation}
where $\pi_{\xAgent[ego]\xAgent[1]}(s',v'_{\xAgent[ego]},v'_{\xAgent[adv]})\pi_{\xAgent[adv]\xAgent[1]}(s',v'_{\xAgent[ego]},v'_{\xAgent[adv]})$ denotes the probabilities of possible combined actions and $q_{\xAgent[2]\xAgent[1]}(s',v'_{\xAgent[ego]},v'_{\xAgent[adv]})$ denotes the Q-values of combined actions at state $(s',v'_{\xAgent[ego]},v'_{\xAgent[adv]})$. We note that the product of $\pi_{\xAgent[ego]\xAgent[1]}(s',v'_{\xAgent[ego]},v'_{\xAgent[adv]})\pi_{\xAgent[adv]\xAgent[1]}(s',v'_{\xAgent[ego]},v'_{\xAgent[adv]})$ and $q_{\xAgent[2]\xAgent[1]}(s',v'_{\xAgent[ego]},v'_{\xAgent[adv]})$ is a scalar.
The Q-function $q_{\xAgent[2]\xAgent[1]}$ for agent $\xAgent[2]$ estimated by agent $\xAgent[1]$, ($\xAgent[1],\xAgent[2] \in\{\xAgent[ego],\xAgent[adv]\}$), is updated as follows (\cref{algLine:updateQ1,algLine:updateQ2}). 
\begin{equation}
    q_{\xAgent[2]\xAgent[1]}(s, v_{\xAgent[ego]}, v_{\xAgent[adv]}, \sgCommonAction_{\xAgent[ego]}, \sgCommonAction_{\xAgent[adv]})
    =
    (1-\alpha_{\xTime[]})q_{\xAgent[2]\xAgent[1]}(s, v_{\xAgent[ego]}, v_{\xAgent[adv]}, \sgCommonAction_{\xAgent[ego]}, \sgCommonAction_{\xAgent[adv]})
    + \alpha_{\xTime[]}(r_{\xAgent[1],\xTime[]} + \gamma \overline{q}_{\xAgent[2]\xAgent[1]}(s', v'_{\xAgent[ego]}, v'_{\xAgent[adv]}))
    \label{eq:updateQ}
\end{equation}
where $\alpha_{\xTime[]}$ is the learning rate at time step $\xTime[]$.

\begin{figure*}
	\centering
		\includegraphics[scale=0.4]{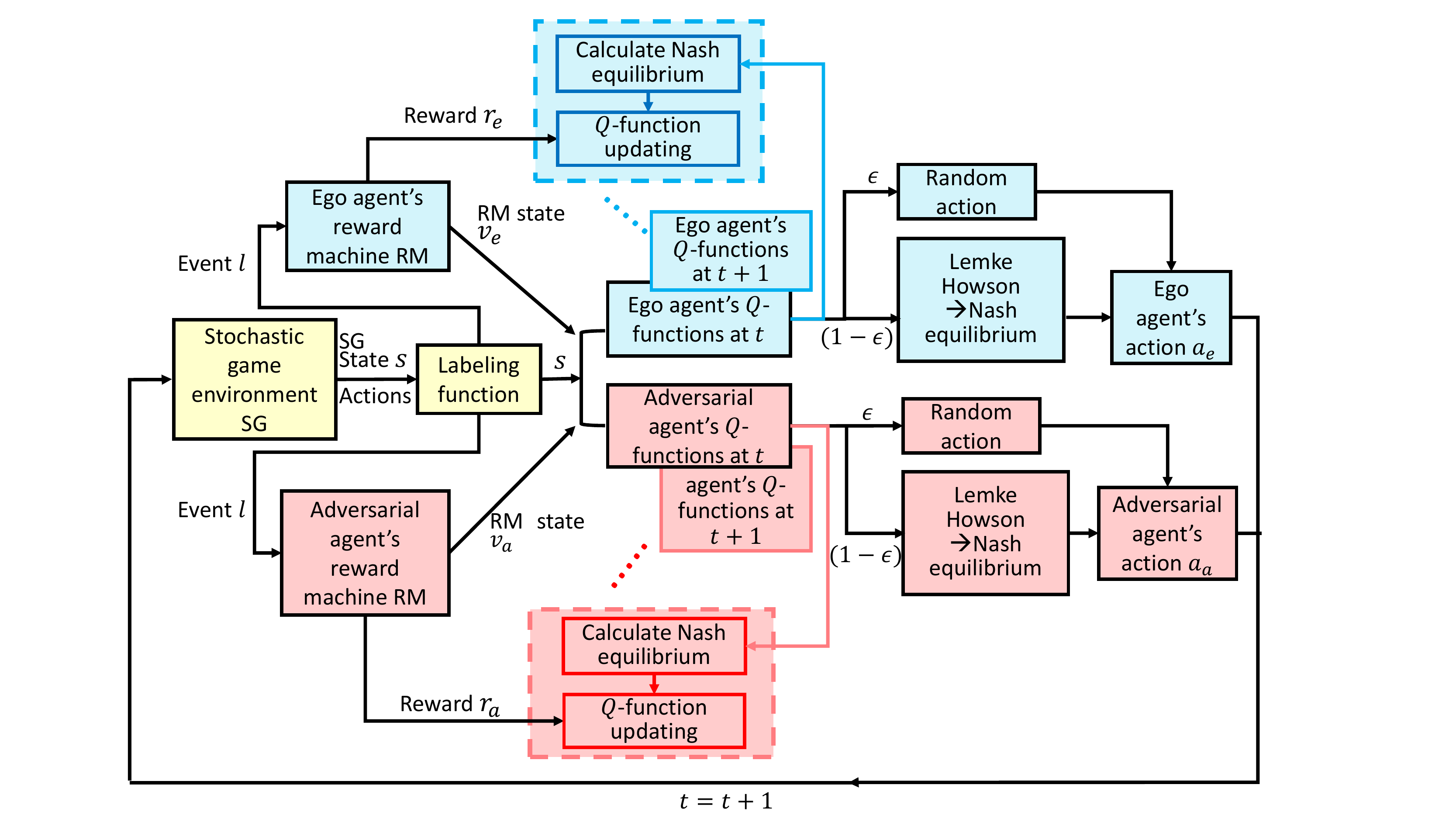}
		\caption{Flowchart of {\algoName}. Following the motivational example, blue indicates the elements related to the ego agent and red indicates the elements related to the adversarial agent. } \label{fig:flowchart}
\end{figure*}

\Cref{fig:flowchart} shows the structure of {\algoName} and focuses on the procedures in one time step. 

{\algoName} is guaranteed to converge to best-response strategies in the limit, as stated in the following theorem and proven in the supplementary materials.

\begin{assumption}
Every state $s\in S$, RM states $v_{\xAgent[ego]} \in V_{\xAgent[ego]},v_{\xAgent[adv]} \in V_{\xAgent[adv]}$, and actions $\sgCommonAction_{\xAgent[ego]} \in A_{\xAgent[ego]}, \sgCommonAction_{\xAgent[adv]} \in A_{\xAgent[adv]}$, are visited infinitely often when the number of episodes goes to infinity.
\end{assumption}

\begin{assumption}
The learning rate $\alpha_{\xTime[]}$ satisfies the following conditions for all $\xTime[]$:
\begin{enumerate}
\item $0 < \alpha_{\xTime[]} < 1, 
\sum_{\xTime[]=\xTime{0}}^{\xTime{\infty}} \alpha_{\xTime[]}(s,v_{\xAgent[ego]},v_{\xAgent[adv]},\sgCommonAction_{\xAgent[ego]},\sgCommonAction_{\xAgent[adv]}) = \infty,
\sum_{\xTime[]=\xTime{0}}^{\xTime{\infty}} [\alpha_{\xTime[]}(s,v_{\xAgent[ego]},v_{\xAgent[adv]},\sgCommonAction_{\xAgent[ego]},\sgCommonAction_{\xAgent[adv]})]^2 < \infty$, and the latter
two hold uniformly and with probability 1.
\item $\alpha_{\xTime[]}(s,v_{\xAgent[ego]},v_{\xAgent[adv]},\sgCommonAction_{\xAgent[ego]},\sgCommonAction_{\xAgent[adv]}) = 0$ if $(s,v_{\xAgent[ego]},v_{\xAgent[adv]},\sgCommonAction_{\xAgent[ego]},\sgCommonAction_{\xAgent[adv]}) \neq (s^{\xTime[]},v_{\xAgent[ego]}^{\xTime[]},v_{\xAgent[adv]}^{\xTime[]},\sgCommonAction_{\xAgent[ego]}^{\xTime[]},\sgCommonAction_{\xAgent[adv]}^{\xTime[]}).$
\end{enumerate}
\end{assumption}

Assumptions 1 and 2 are standard assumptions and similar to those in Q-learning \cite{melo2001convergence}. Condition 2 in Assumption 2 requires that at each step, only the Q-function elements related to the current state, RM state, and actions are updated.

\begin{assumption}
One of the following conditions holds during learning.
\begin{enumerate}
    \item
    Every stage game $(q_{\xAgent[ego]\xAgent[1]}(s,v_{\xAgent[ego]},v_{\xAgent[adv]}),q_{\xAgent[adv]\xAgent[1]}(s,v_{\xAgent[ego]},v_{\xAgent[adv]}))$,
    $\xAgent[1] \in \{\xAgent[ego],\xAgent[adv]\}$, for all $\xTime[], s,v_{\xAgent[ego]},$ and $v_{\xAgent[adv]}$, has a global optimal point ($\Tilde{\pi}_{\xAgent[ego]\xAgent[1]}, \Tilde{\pi}_{\xAgent[adv]\xAgent[1]}$), \big(i.e., $\Tilde{\pi}_{\xAgent[2]\xAgent[1]} q_{\xAgent[2]\xAgent[1]} \ge \Tilde{\pi}'_{\xAgent[2]\xAgent[1]} q_{\xAgent[2]\xAgent[1]}$ for any $\Tilde{\pi}'_{\xAgent[2]\xAgent[1]},$  $\xAgent[2] \in \{\xAgent[ego],\xAgent[adv]\}$\big)  and agents’ rewards in this equilibrium are used to update their Q-functions.
    \item
    Every stage game ($q_{\xAgent[ego]\xAgent[1]}(s,v_{\xAgent[ego]},v_{\xAgent[adv]}),q_{\xAgent[adv]\xAgent[1]}(s,v_{\xAgent[ego]},v_{\xAgent[adv]})$), $\xAgent[1] \in \{\xAgent[ego],\xAgent[adv]\}$, for all $\xTime[], s,v_{\xAgent[ego]},$ and $v_{\xAgent[adv]}$, has a saddle point ($\Tilde{\pi}_{\xAgent[ego]\xAgent[1]}, 
    \Tilde{\pi}_{\xAgent[adv]\xAgent[1]}$), \big(i.e., $\Tilde{\pi}_{\xAgent[2]\xAgent[1]} \Tilde{\pi}_{-\xAgent[2]\xAgent[1]}q_{\xAgent[2]\xAgent[1]} \ge \Tilde{\pi}'_{\xAgent[2]\xAgent[1]} \Tilde{\pi}_{-\xAgent[2]\xAgent[1]}q_{\xAgent[2]\xAgent[1]}$ for any $\Tilde{\pi}'_{\xAgent[2]\xAgent[1]} $, $\Tilde{\pi}_{\xAgent[2]\xAgent[1]} \Tilde{\pi}_{-\xAgent[2]\xAgent[1]}q_{\xAgent[2]\xAgent[1]} \le \Tilde{\pi}_{\xAgent[2]\xAgent[1]} \Tilde{\pi}'_{-\xAgent[2]\xAgent[1]}q_{\xAgent[2]\xAgent[1]}$ for any $\Tilde{\pi}'_{-\xAgent[2]\xAgent[1]}$, $\xAgent[2] \in \{\xAgent[ego],\xAgent[adv]\}$\big), and agents’ rewards in this equilibrium are used to update their Q-functions.
\end{enumerate}
\end{assumption}

In Assumption 3, $\Tilde{\pi}_{\xAgent[2]\xAgent[1]}$ denotes the strategy of agent \xAgent[2] at either the global optimal point or the saddle point, given the Q-functions learned by agent \xAgent[1]. $\Tilde{\pi}_{-\xAgent[2]\xAgent[1]}$ is the strategy of the agent other than agent $\xAgent[2]$, i.e., $\Tilde{\pi}_{-\xAgent[ego]\xAgent[1]} = \Tilde{\pi}_{\xAgent[adv]\xAgent[1]}, \Tilde{\pi}_{-\xAgent[adv]\xAgent[1]} = \Tilde{\pi}_{\xAgent[ego]\xAgent[1]}.$ Assumption 3 requires that the stage game at each time step has either a global optimal point or a saddle point. 

\begin{thm}
Under Assumptions 1, 2 and 3,  the sequence $q_{\xAgent[1]\xTime[]} = (q_{\xAgent[ego]\xAgent[1]}^{\xTime[]}, q_{\xAgent[adv]\xAgent[1]}^{\xTime[]})$ at time $t$, $\xAgent[1] \in \{\xAgent[ego],\xAgent[adv]\}$ updated by\\ 
\begin{equation}
        q_{\xAgent[2]\xAgent[1]}^{\xTime[]+1}(s, v_{\xAgent[ego]}, v_{\xAgent[adv]}, \sgCommonAction_{\xAgent[ego]}, \sgCommonAction_{\xAgent[adv]})
        =
        (1-\alpha_{\xTime[]})q_{\xAgent[2]\xAgent[1]}^{\xTime[]}(s, v_{\xAgent[ego]}, v_{\xAgent[adv]}, \sgCommonAction_{\xAgent[ego]}, \sgCommonAction_{\xAgent[adv]})
        + \alpha_{\xTime[]}\big(r_{\xAgent[2],\xTime[]} + \gamma \pi_{\xAgent[ego]\xAgent[1]}(\cdot|s', v'_{\xAgent[ego]}, v'_{\xAgent[adv]})\pi_{\xAgent[adv]\xAgent[1]}(\cdot|s', v'_{\xAgent[ego]}, v'_{\xAgent[adv]})q_{\xAgent[2]\xAgent[1]}^{\xTime[]}(s',v'_{\xAgent[ego]}, v'_{\xAgent[adv]})\big)
\end{equation}
for $\xAgent[2] \in \{\xAgent[ego],\xAgent[adv]\}$, where $(\pi_{\xAgent[ego]\xAgent[1]}(\cdot|s', v'_{\xAgent[ego]}, v'_{\xAgent[adv]}),\pi_{\xAgent[adv]\xAgent[1]}(\cdot|s', v'_{\xAgent[ego]}, v'_{\xAgent[adv]}))$ is the appropriate type of Nash equilibrium solution for the stage game $(q_{\xAgent[ego]\xAgent[1]}^{\xTime[]}(s', v'_{\xAgent[ego]}, v'_{\xAgent[adv]}), q_{\xAgent[adv]\xAgent[1]}^{\xTime[]}(s', v'_{\xAgent[ego]}, v'_{\xAgent[adv]}))$, converges to the Q-functions at a Nash equilibrium 
$(q^{\ast}_{\xAgent[ego]\xAgent[1]}, q^{\ast}_{\xAgent[adv]\xAgent[1]})$.
\end{thm}

Note that $(\pi_{\xAgent[ego]\xAgent[1]}(\cdot|s', v'_{\xAgent[ego]}, v'_{\xAgent[adv]}),\pi_{\xAgent[adv]\xAgent[1]}(\cdot|s', v'_{\xAgent[ego]}, v'_{\xAgent[adv]}))$ is the appropriate type of Nash equilibrium solution if the strategy corresponds to the global optimal point or the saddle point for the stage game $(q_{\xAgent[ego]\xAgent[1]}^{\xTime[]}(s', v'_{\xAgent[ego]}, v'_{\xAgent[adv]}), q_{\xAgent[adv]\xAgent[1]}^{\xTime[]}(s', v'_{\xAgent[ego]}, v'_{\xAgent[adv]}))$. $q^{\ast}_{\xAgent[1]\xAgent[2]}$ is the Q-function at a Nash equilibrium for agent $\xAgent[1]$ and learned by agent $\xAgent[2]$.

%% file: experiment.tex
\section{Experiments}
\label{sec_experiment}

In this section, we evaluate the effectiveness of the proposed {\algoName} method in three case studies. We compare {\algoName} with following baseline methods:
\begin{itemize}
 \item Nash-Q: we perform Nash Q-learning algorithm developed in~\cite{hu2003nash} and the agents' locations are taken as the state. \\
 \item Nash-QAS (Nash Q-learning in augmented state space): to have more information on high-level events, we include an extra binary vector in the state representing whether each event has been encountered or not.\\
 \item MADDPG-SG (multi-agent deep deterministic policy gradient with state of the stochastic game): we adopt one of the state-of-the-art MARL baselines --- MADDPG algorithm developed in~\cite{lowe2017multi} --- and include the agents' locations as the state.\\
 \item MADDPG-AS (multi-agent deep deterministic policy gradient in augmented state space): similar to Nash-QAS, we include an extra binary vector in the state to represent high-level information for MADDPG~\cite{lowe2017multi}.

\end{itemize}

In each case study, each agent is given a task with sparse rewards to achieve, which can be specified as a reward machine. Each agent receives a reward of one if and only if the task is completed. We have two agents in a 6$\times$6 grid world for three case studies, as shown in \Cref{fig:grid_map}. Following the notations in the motivational example, we use the \colorname[Ego] color to indicate the locations corresponding to the ego agent and the \colorname[Adv] color for the adversarial agent. The power bases are represented by circles and starting locations are represented by triangles. At each time step, each agent can select from: \{up, down, left, right\}. Each action has a slip rate of 0.5\% and $\epsilon$ is set to 0.25. 
\begin{figure}
		\centering
		\input{figs/tasks-map2.tikz}
		\caption{
		    The grid world for case studies.
		    In Case Study I and II, the adversarial agent starts from location `a'.
		    In Case Study III, the adversarial agent starts from location `a' or `b'. } \label{fig:grid_map}
\end{figure}
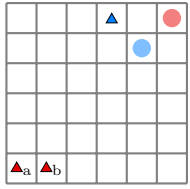

\begin{figure}
    \centering
    
    \begin{subfigure}[t]{.5\linewidth}
        \centering
        \input{figs/tasks-rm1.tikz}%
        \caption{Case study I.}
        \label{fig:tasks-rm:cs1}
    \end{subfigure}%
    \begin{subfigure}[t]{.5\linewidth}
        \centering
        \input{figs/tasks-rm2.tikz}%
        \caption{Case studies II and III.}
        \label{fig:tasks-rm:cs23}
    \end{subfigure}%

    
    \caption{
        The reward machines used in the case studies.
        In each case study, the reward machine of each agent has the same structure, only the reward differs.
        Hence, we present the rewards for agent \protect\agent[Ego] in \colorname[Ego], and the rewards for agent \protect\agent[Adv] in \colorname[Adv].
        The reward is sparse (zero for both agents when not specified on a transition).
        Omitted transitions are self-looping transitions with reward 0. 
    }
    \label{fig:tasks-rm}
\end{figure}
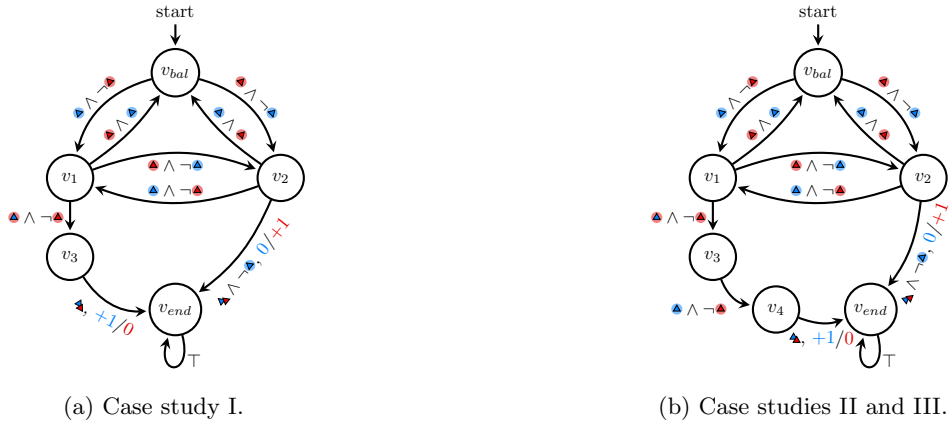

\begin{figure} 
    \centering
    
    \begin{subfigure}[hbt!]{.45\linewidth} 
        \centering
        \includegraphics[width=\linewidth]{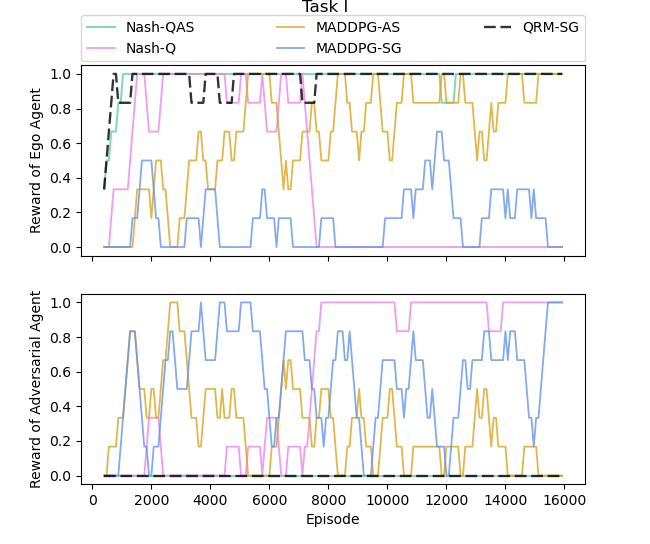}
        \caption{Case study I.}
        \label{fig:task1_R}
    \end{subfigure} \\ %
    \begin{subfigure}[hbt!]{.45\linewidth}
        \centering
        \includegraphics[width=\linewidth]{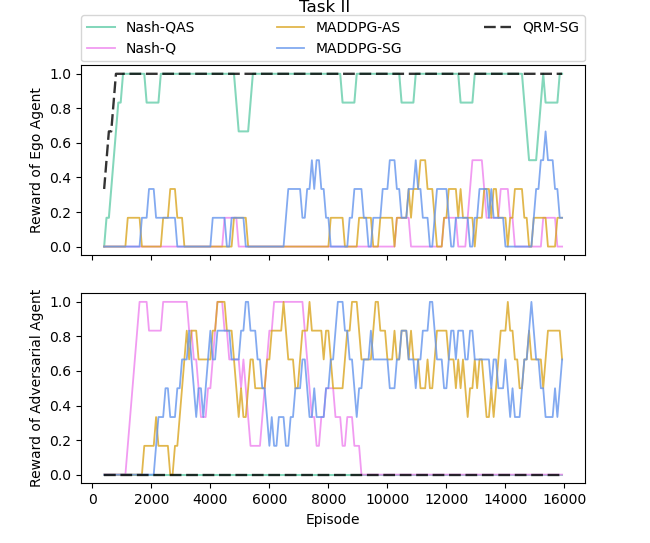}
        \caption{Case study II.}
        \label{fig:task2_R}
    \end{subfigure} \\ %
    \begin{subfigure}[hbt!]{.45\linewidth}
        \centering
        \includegraphics[width=\linewidth]{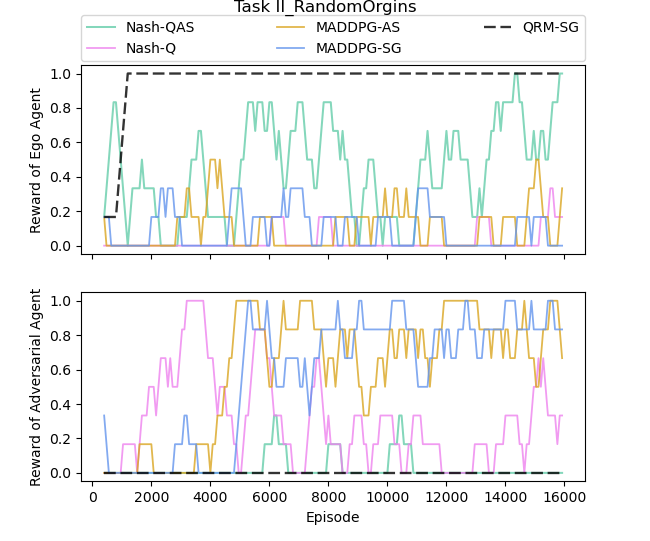}
        \caption{Case study III.}
        \label{fig:task3_R}
    \end{subfigure}%
    
    \caption{
        Cumulative reward comparison for each case study. 
        Smoothed plot with a rolling window of size 6. {\algoName} is the proposed method.
    }
    \label{fig:tasks_R}
\end{figure}

The first case study is designed as an 'appetizer' to test {\algoName}. Then, we make the ego agent's task more demanding in the second and third case studies. Randomnesses in the starting locations of the adversarial agent are included in the third case study. We define that one agent is captured by the other agent when the distance between two agents is smaller than 2. Each agent is considered to complete the task after capturing the other agent. Only when an agent is the more powerful of the two agents, it has the capability to capture the other agent. We set different conditions for the ego agent to be the more powerful in different case studies. The tasks for agents to complete are specified as reward machines, which are demonstrated in \Cref{fig:tasks-rm}. The adversarial agent is given the same task in three case studies, which requires it to be the more powerful by reaching its power base (\signal[AdvAtHome]) and then capture the ego agent (\signal[EgoMeetAdv]). In Case Study I, the ego agent is required to first reach its own power base (\signal[EgoAtHome]), then destroy/reach the adversarial agent's power base (\signal[EgoAtAdvHome]) to be the more powerful, and capture the adversarial agent afterward (\signal[EgoMeetAdv]). In Case Study II, the required sequential events for the ego agent to be powerful are: reaching its power base (\signal[EgoAtHome]), reaching the adversarial agent's power base (\signal[EgoAtAdvHome]), reaching its power base (\signal[EgoAtHome]). These events demonstrate the scenario in that the ego agent first gets energy at its power base, destroys the adversarial agent's power base using most of its energy, and then gets recharged to capture the adversarial agent. Case Study III is different from Case Study II in that the adversarial agent randomly samples the starting location from 2 possible locations. In all case studies, we define that after the adversarial agent arrives at its power base, the ego agent has to reach its own power base again to be able to destroy the adversarial agent's power base. After a power base is destroyed, the corresponding agent is not possible to be more powerful.

To better analyze the results, each case study is designed to have best-response strategies for agents. The ego agent is expected to complete the required sequence of events to be the more powerful agent, capture the adversarial agent, and complete the task resulting in a cumulative reward of 1. As the adversarial agent is far away from its power base, its power base would be destroyed by the ego agent before it arrives at a Nash equilibrium. Thus, the adversarial agent would fail to complete its task. The learning processes of each agent in three case studies are plotted in \Cref{fig:task1_R,fig:task2_R,fig:task3_R}. Every 80 episodes, we stop learning, test the algorithms' performance, and save the cumulative rewards of each agent. 
In Case Study I, {\algoName} finds the Nash equilibrium in around 7500 episodes. Nash-QAS finds the Nash equilibrium after 12000 episodes, which indicates that using the augmented state can be sufficient for the ego agent to learn to complete the task. When learning by Nash-Q, the adversarial agent completes the task. The reason can be that the task for the adversarial agent is much easier. MADDPG-SG and MADDPG-AS do not converge within 16000 episodes. 
In MADDPG-SG, the ego agent needs more episodes for completing the task compared to MADDPG-AS.
One possible reason is that MADDPG-AS perceives more information represented in the augmented state space.
In Case Study II, the policies of agents reach the Nash equilibrium after around 1000 episodes utilizing {\algoName}, while baselines fail to converge to the Nash equilibrium. 
In Case Study III, {\algoName} finds the Nash equilibrium in around 1500 episodes, while baselines have difficulty converging to the Nash equilibrium. The ego agent learned by Nash-QAS receives a higher reward than other baselines. The ego agent learned by other baselines rarely finishes the task. 
From the results of the three case studies, {\algoName} outperforms the four baseline methods, where the ego agent can accomplish the task at a Nash equilibrium in several thousand episodes.

An analysis of the effect of $\epsilon$ on the performance of QRM-SG is available in the supplementary materials. Additionally, an extensive evaluation of QRM-SG on a 12$\times$12 grid world is conducted in the supplementary materials.




%% file: figs/tasks-map2.tikz
\begin{tikzpicture}[
    scale=0.8,
    thick,
    every node/.append style={transform shape},
]

\begin{scope}[
    xscale=.5,
    yscale=.5, 
    every node/.append style={scale=2}, 
]

\draw[step=1,color=gray,shift={(-.5,-.5)}] (0,0) grid +(6,6);

\node at (4,4) {\Large\location[EgoHome]};
\node at (5,5) {\Large\location[AdvHome]};
\node at (3,5) {\agent[Ego]};
\node at (0,0) {\agent[Adv]\textsubscript{a}};
\node at (1,0) {\agent[Adv]\textsubscript{b}};

\end{scope}

\end{tikzpicture}

%% file: figs/tasks-rm1.tikz
\begin{tikzpicture}[
    scale=.7,
    thick,
    every node/.append style={transform shape},
]

\node[state,initial above] (uBal) 
    at ( 0,-0)
    {$\mealyCommonState_{bal}$};
\node[state] (uEgo)
    at (-2.0,-2.0)
    {$\mealyCommonState_{\textcolor{black}{1}}$};
\node[state] (uAdv)
    at (+2.0,-2.0)
    {$\mealyCommonState_{\textcolor{black}{2}}$};
\node[state] (uEgo2)
    at (-2.0,-3.5)
    {$\mealyCommonState_{\textcolor{black}{3}}$};
\node[state] (uEnd)
    at ( 0.0,-4.5)
    {$\mealyCommonState_{end}$};

\path[->,sloped]

(uBal)
edge[bend right] node[]
    {$\signal[EgoAtHome]\land\lnot\signal[AdvAtHome]$}
    (uEgo)
edge[bend left] node[]
    {$\signal[AdvAtHome]\land\lnot\signal[EgoAtHome]$}
    (uAdv)
    
(uEgo)
edge[bend right=10] node[]
    {$\signal[AdvAtHome]\land\signal[EgoAtHome]$}
    (uBal)
edge[bend left=20] node[swap]
    {$\signal[AdvAtHome]\land\lnot\signal[EgoAtHome]$}
    (uAdv)
edge[] node[swap,sloped=false]
    {$\signal[EgoAtAdvHome]\land\lnot\signal[AdvAtHome]$}
    (uEgo2)

(uAdv)
edge[bend left=10] node[]
    {$\signal[EgoAtHome]\land\signal[AdvAtHome]$}
    (uBal)
edge[bend left=20] node[]
    {$\signal[EgoAtHome]\land\lnot\signal[AdvAtHome]$}
    (uEgo)
edge[bend left=15] node[swap]
    {$\signal[EgoMeetAdv]\land\lnot\signal[EgoAtHome]$,
    \textcolor{colorPlayerEgo}{$0$}/\textcolor{colorPlayerAdv}{$+1$}}
    (uEnd)

(uEgo2)
edge[bend right] node[swap]
    {$\signal[EgoMeetAdv]$,
    \textcolor{colorPlayerEgo}{$+1$}/\textcolor{colorPlayerAdv}{$0$}}
    (uEnd)

(uEnd)
edge[out=-75,in=-105,loop] node[sloped=false,xshift=+1em, yshift=1em]   {$\top$} 
(uEnd)
;

\end{tikzpicture}

%% file: figs/tasks-rm2.tikz
\begin{tikzpicture}[
    scale=.7,
    thick,
    every node/.append style={transform shape},
]
\node[state,initial above] (uBal) 
    at ( 0,-0)
    {$\mealyCommonState_{bal}$};
\node[state] (uEgo)
    at (-2.0,-2.0)
    {$\mealyCommonState_{\textcolor{black}{1}}$};
\node[state] (uAdv)
    at (+2.0,-2.0)
    {$\mealyCommonState_{\textcolor{black}{2}}$};
\node[state] (uEgo2)
    at (-2.0,-3.5)
    {$\mealyCommonState_{\textcolor{black}{3}}$};
\node[state] (uEgo3)
    at (-0.8,-4.5)
    {$\mealyCommonState_{\textcolor{black}{4}}$};
\node[state] (uEnd)
    at (+1.0,-4.5)
    {$\mealyCommonState_{end}$};

\path[->,sloped]

(uBal)
edge[bend right] node[]
    {$\signal[EgoAtHome]\land\lnot\signal[AdvAtHome]$}
    (uEgo)
edge[bend left] node[]
    {$\signal[AdvAtHome]\land\lnot\signal[EgoAtHome]$}
    (uAdv)
    
(uEgo)
edge[bend right=10] node[]
    {$\signal[AdvAtHome]\land\signal[EgoAtHome]$}
    (uBal)
edge[bend left=20] node[swap]
    {$\signal[AdvAtHome]\land\lnot\signal[EgoAtHome]$}
    (uAdv)
edge[] node[swap,sloped=false]
    {$\signal[EgoAtAdvHome]\land\lnot\signal[AdvAtHome]$}
    (uEgo2)

(uAdv)
edge[bend left=10] node[]
    {$\signal[EgoAtHome]\land\signal[AdvAtHome]$}
    (uBal)
edge[bend left=20] node[]
    {$\signal[EgoAtHome]\land\lnot\signal[AdvAtHome]$}
    (uEgo)
edge[bend left=15] node[swap]
    {$\signal[EgoMeetAdv]\land\lnot\signal[EgoAtHome]$,
    \textcolor{colorPlayerEgo}{$0$}/\textcolor{colorPlayerAdv}{$+1$}}
    (uEnd)

(uEgo2)
edge[bend right] node[swap,sloped=false]
    {$\signal[EgoAtHome]\land\lnot\signal[AdvAtHome]$}
    (uEgo3)

(uEgo3)
edge[bend right=20] node[swap]
    {$\signal[EgoMeetAdv]$,
    \textcolor{colorPlayerEgo}{$+1$}/\textcolor{colorPlayerAdv}{$0$}}
    (uEnd)

(uEnd)
edge[out=-75,in=-105,loop] node[sloped=false,xshift=+1em, yshift=1em]   {$\top$} 
(uEnd)
;

\end{tikzpicture}

%% file: conclusion.tex
\section{Conclusions}
In this paper, we introduce the utilization of reward machines to expose the structure of non-Markovian reward functions for the learning agents in two-agent general-sum stochastic games. Each task is specified by a reward machine. We propose {\algoName{}} as a variant of Q-learning for the setting of stochastic games and integrated with reward machines to learn best-response strategies at a Nash equilibrium for each agent. We prove that {\algoName{}} converges to Q-functions at a Nash equilibrium under certain conditions. Three case studies are conducted to evaluate the performance of {\algoName{}}.

This paper opens the door for using reward machines in stochastic games. First, one immediate extension is to jointly learn reward machines and best-response strategies during RL. Second, extending the methodology to general-sum stochastic games with more agents is worth further investigation. Finally, the same methodology can be readily applied to other forms of RL, such as model-based RL, or actor-critic methods.

%% file: appendix.tex
\appendix

\begin{align*}\nonumber
\centering
\scalebox{1.5}{\textrm{Reinforcement Learning}}&\scalebox{1.5}{\textrm{ With Reward Machines in Stochastic Games}} \\
&\scalebox{1.5}{\textrm{Supplementary Materials}}
\end{align*}


\hfill \break

\setcounter{lemma}{1}
\setcounter{definition}{6}
\addcontentsline{toc}{section}{Appendices}
\renewcommand{\thesubsection}{\Alph{subsection}}

\section{Proof of Theorem 1}

In the supplementary material, we provide the convergence proof of QRM-SG in Theorem 1. Recall that agent $\xAgent[1]$ ($\xAgent[1] \in \{\xAgent[ego],\xAgent[adv]\}$) in {\algoName} maintains the Q-functions of both agents $q_{\xAgent[1]} = (q_{\xAgent[ego]\xAgent[1]}, q_{\xAgent[adv]\xAgent[1]})$ in the two-agent stochastic game. Agents aim to find the best-response strategies given their own estimations of the Q-functions of both agents. We prove that $(q_{\xAgent[ego]\xAgent[1]}, q_{\xAgent[adv]\xAgent[1]})$ converges to the Q-functions at a Nash equilibrium $(q^\ast_{\xAgent[ego]}, q^\ast_{\xAgent[adv]})$ for agent $\xAgent[1]$.
In the supplementary materials, to simplify the notation, we omit the subscript ${\xAgent[1]}$ (which represents the learning agent $\xAgent[1]$) in the Q-functions. Each agent is equipped with $q=(q_{\xAgent[ego]}, q_{\xAgent[adv]})$ and strategy estimations of both agents $\pi=(\pi_{\xAgent[ego]}, \pi_{\xAgent[adv]})$.
The proof is based on the following lemma in 
\cite{szepesvari1999unified}.

\begin{lemma} \label{lemma1:follow}
Assume that $\alpha_t$ satisfies Assumption 2 and the mapping $F^t: \mathbb{Q} \rightarrow \mathbb{Q}$ satisfies the following condition: there exists a number $0<\beta<1$ and a sequence $\lambda^t \ge 0$ converging to zero with probability 1 such that $ || F^t q - F^t q^{\ast} || \le \beta ||q-q^{\ast}|| + \lambda^t$
for all $q \in \mathbb{Q}$ and $q^{\ast} = \mathbb{E}[F^t q^{\ast}]$, then the iteration defined by
\begin{equation}
\label{lemma:follow}
    q^{t+1} = (1-\alpha_t)q^t + \alpha_t[F^tq^t]
\end{equation}
converges to $q^{\ast}$ with probability 1.
\end{lemma}

Q-functions in {\algoName} uses the same updating structure as the iteration equation in \Cref{lemma:follow}. Thus, we show the convergence of Q-functions in {\algoName} to the Q-functions at a Nash equilibrium by adapting \Cref{lemma:follow} to {\algoName}. We first define the operator $F^t$ in the two-agent stochastic game and then prove $F^t$ satisfies the condition in \Cref{lemma1:follow}.

For each agent, we have $q=(q_{\xAgent[ego]}, q_{\xAgent[adv]})$, where $q \in \mathbb{Q} \text{ and } q_{\xAgent[2]} \in \mathbb{Q}_{\xAgent[2]}$ for $\xAgent[2]={\xAgent[ego]}, {\xAgent[adv]}$. Thus, $\mathbb{Q}=\mathbb{Q}_{\xAgent[ego]}\times \mathbb{Q}_{\xAgent[adv]}$. Recall that the Q-functions in {\algoName} are in augmented state space, taking the state of the stochastic game and states of the reward machines as input. 
\begin{definition}
\label{def:map}
We define the mapping $F^t: \mathbb{Q} \rightarrow \mathbb{Q}$ is on the complete metric space $\mathbb{Q}$ to $\mathbb{Q}$, $F^t q = (F^tq_{\xAgent[ego]}, F^tq_{\xAgent[adv]})$, where
\begin{equation}
\begin{aligned}
    \text{For } \xAgent[2] \in \{\xAgent[ego],\xAgent[adv]\}, F^tq_{\xAgent[2]}(s,v_{\xAgent[ego]},v_{\xAgent[adv]},a_{\xAgent[ego]},a_{\xAgent[adv]}) = r_{\xAgent[2],\xTime[]} +     \gamma \pi_{\xAgent[ego]}(s', v'_{\xAgent[ego]}, v'_{\xAgent[adv]})\pi_{\xAgent[adv]}(s', v'_{\xAgent[ego]}, v'_{\xAgent[adv]})q_{\xAgent[2]}(s',v'_{\xAgent[ego]}, v'_{\xAgent[adv]}) 
\end{aligned}
\end{equation}
where $(s',v'_{\xAgent[ego]},v'_{\xAgent[adv]})$ represents the state and RM states in the next time step, and $(\pi_{\xAgent[ego]}(s', v'_{\xAgent[ego]}, v'_{\xAgent[adv]}), \pi_{\xAgent[adv]}(s', v'_{\xAgent[ego]}, v'_{\xAgent[adv]}))$ represents a Nash equilibrium solution for the stage game ($q_{\xAgent[ego]}(s',v'_{\xAgent[ego]},v'_{\xAgent[adv]}),q_{\xAgent[adv]}(s',v'_{\xAgent[ego]},v'_{\xAgent[adv]})$). $\pi_{\xAgent[ego]}(s',v'_{\xAgent[ego]},v'_{\xAgent[adv]})\pi_{\xAgent[adv]}(s',v'_{\xAgent[ego]},v'_{\xAgent[adv]})$ denotes the probabilities of all possible combined actions and $q_{\xAgent[2]}(s',v'_{\xAgent[ego]},v'_{\xAgent[adv]})$ denotes all Q-values of combined actions at state $(s',v'_{\xAgent[ego]},v'_{\xAgent[adv]})$. We note that the product of $\pi_{\xAgent[ego]}(s',v'_{\xAgent[ego]},v'_{\xAgent[adv]})\pi_{\xAgent[adv]}(s',v'_{\xAgent[ego]},v'_{\xAgent[adv]})$ and $q_{\xAgent[2]}(s',v'_{\xAgent[ego]},v'_{\xAgent[adv]})$ is a scalar.
\end{definition}

Recall that in an SG, agent $\xAgent[2]$ $(\xAgent[2] \in \{\xAgent[ego],\xAgent[adv]\})$ aims to maximize the expected discounted cumulative reward $\Tilde{v}_{\xAgent[2]}$. In the stage game ($q^{\ast}_{\xAgent[ego]}(s,v_{\xAgent[ego]},v_{\xAgent[adv]}), q^{\ast}_{\xAgent[adv]}(s,v_{\xAgent[ego]},v_{\xAgent[adv]})$), the optimal discounted cumulative reward $\Tilde{v}_{\xAgent[2]}$ can be linked to the rewards when the agents reach a Nash equilibrium based on the following lemma in 
\cite{filar2012competitive}.

\begin{lemma}
The following assertions are equivalent:
\begin{enumerate}
\label{lemma:qandv}
    \item 
    $(\pi^{\ast}_{\xAgent[ego]}, \pi^{\ast}_{\xAgent[adv]})$ is a Nash equilibrium point in a stochastic game with rewards \big($\Tilde{v}_{\xAgent[ego]}(\pi^{\ast}_{\xAgent[ego]}, \pi^{\ast}_{\xAgent[adv]}),\Tilde{v}_{\xAgent[adv]}(\pi^{\ast}_{\xAgent[ego]}, \pi^{\ast}_{\xAgent[adv]})$\big), where $\Tilde{ v}_{\xAgent[2]}(\pi^{\ast}_{\xAgent[ego]}, \pi^{\ast}_{\xAgent[adv]})= \big(\Tilde{v}_{\xAgent[2]}(s^1,v^1_{\xAgent[ego]},v^1_{\xAgent[adv]},\pi^{\ast}_{\xAgent[ego]}, \pi^{\ast}_{\xAgent[adv]}),\dots,\Tilde{v}_{\xAgent[2]}(s^k, v^k_{\xAgent[ego]},v^k_{\xAgent[adv]}, \pi^{\ast}_{\xAgent[ego]}, \pi^{\ast}_{\xAgent[adv]})\big), \xAgent[2] \in \{\xAgent[ego],\xAgent[adv]\}$
    \item 
    For each $(s,v_{\xAgent[ego]},v_{\xAgent[adv]}) \in S\times V_{\xAgent[ego]} \times V_{\xAgent[adv]}$, a Nash equilibrium point $\big(\pi^{\ast}_{\xAgent[ego]}(s,v_{\xAgent[ego]},v_{\xAgent[adv]}), \pi^{\ast}_{\xAgent[adv]}(s,v_{\xAgent[ego]},v_{\xAgent[adv]})\big)$ in the stage game $\big(q^{\ast}_{\xAgent[ego]}(s,v_{\xAgent[ego]},v_{\xAgent[adv]}), q^{\ast}_{\xAgent[adv]}(s,v_{\xAgent[ego]},v_{\xAgent[adv]})\big)$ achieves rewards $\big(\Tilde{v}_{\xAgent[ego]}(s,v_{\xAgent[ego]},v_{\xAgent[adv]},\pi^{\ast}_{\xAgent[ego]},\pi^{\ast}_{\xAgent[adv]}),\Tilde{v}_{\xAgent[adv]}(s,v_{\xAgent[ego]},v_{\xAgent[adv]},\pi^{\ast}_{\xAgent[ego]},\pi^{\ast}_{\xAgent[adv]})  \big)$, where for $\xAgent[2] \in \{\xAgent[ego], \xAgent[adv]\}$,
    \begin{equation}
    \begin{aligned}
        & q^{\ast}_{\xAgent[2]} = r_{\xAgent[2]} +  \gamma \sum_{\substack{s' \in S \\ v'_{\xAgent[ego]} \in V_{\xAgent[ego]} \\ v'_{\xAgent[adv]} \in V_{\xAgent[adv]}}} p(s',v'_{\xAgent[ego]},v'_{\xAgent[adv]}|s,v_{\xAgent[ego]},v_{\xAgent[adv]}, \sgCommonAction_{\xAgent[ego]}, \sgCommonAction_{\xAgent[adv]}) \Tilde{v}_{\xAgent[2]}(s', v'_{\xAgent[ego]},v'_{\xAgent[adv]},\pi_{\xAgent[ego]}^\ast, \pi_{\xAgent[adv]}^\ast)
        \end{aligned}
    \end{equation}
\end{enumerate}
\end{lemma}

\Cref{lemma:qandv} demonstrates the relationship between $q^{\ast}_{\xAgent[2]}$ and $\Tilde{v}_{\xAgent[2]}$,  $\Tilde{v}_{\xAgent[2]}(s,v_{\xAgent[ego]},v_{\xAgent[adv]}) = \pi^{\ast}_{\xAgent[ego]}(s,v_{\xAgent[ego]},v_{\xAgent[adv]})\pi^{\ast}_{\xAgent[adv]}(s,v_{\xAgent[ego]},v_{\xAgent[adv]})q^{\ast}(s,v_{\xAgent[ego]},v_{\xAgent[adv]})$. With \Cref{lemma:qandv}, we can achieve $q^{\ast}=\mathbb{E}[F^tq^{\ast}]$ shown as follows.
\begin{lemma}
\label{lemma:exp}
    For a two-agent stochastic game, given that $F^t$ is defined in \Cref{def:map}, $q^{\ast}$ is the Q-functions at a Nash equilibrium, and $q^{\ast}=(q^{\ast}_{\xAgent[ego]},q^{\ast}_{\xAgent[adv]})$, we have $q^{\ast}=\mathbb{E}[F^tq^{\ast}]$.  
\end{lemma}

\begin{proof}
In the stage game $\big(q^{\ast}_{\xAgent[ego]}(s',v'_{\xAgent[ego]},v'_{\xAgent[adv]}), q^{\ast}_{\xAgent[adv]}(s',v'_{\xAgent[ego]},v'_{\xAgent[adv]})\big)$, given the strategies at a Nash equilibrium point $\big(\pi^{\ast}_{\xAgent[ego]}(s',v'_{\xAgent[ego]},v'_{\xAgent[adv]}), \\ \pi^{\ast}_{\xAgent[adv]}(s',v'_{\xAgent[ego]},v'_{\xAgent[adv]})\big)$ and the reward $\Tilde{v}_{\xAgent[2]}(s',v'_{\xAgent[ego]},v'_{\xAgent[adv]}, \pi^{\ast}_{\xAgent[ego]},\pi^{\ast}_{\xAgent[adv]})$ of agent $\xAgent[2]$ obtained at the Nash equilibrium point, $\Tilde{v}_{\xAgent[2]}(s',v'_{\xAgent[ego]},v'_{\xAgent[adv]},\pi^{\ast}_{\xAgent[ego]},\pi^{\ast}_{\xAgent[adv]}) \\ = \pi^{\ast}_{\xAgent[ego]}(s',v'_{\xAgent[ego]},v'_{\xAgent[adv]})  \pi^{\ast}_{\xAgent[adv]}(s',v'_{\xAgent[ego]},v'_{\xAgent[adv]}) q^{\ast}_{\xAgent[2]}(s',v'_{\xAgent[ego]},v'_{\xAgent[adv]})$.  
The reward at a Nash equilibrium point depends on the best-response strategies of both agents. 
Following \Cref{lemma:qandv}, for $\xAgent[2] \in \{\xAgent[ego], \xAgent[adv]\}$,

\begin{equation}
    \begin{aligned}
        q^{\ast}_{\xAgent[2]} &= r_{\xAgent[2]} + \allowbreak
         \gamma \sum_{\substack{s' \in S \\ v'_{\xAgent[ego]} \in V_{\xAgent[ego]} \\ v'_{\xAgent[adv]} \in V_{\xAgent[adv]}}} p(s',v'_{\xAgent[ego]},v'_{\xAgent[adv]}|s,v_{\xAgent[ego]},v_{\xAgent[adv]}, \sgCommonAction_{\xAgent[ego]}, \sgCommonAction_{\xAgent[adv]}) \Tilde{v}_{\xAgent[2]}(s', v'_{\xAgent[ego]},v'_{\xAgent[adv]},\pi_{\xAgent[ego]}^\ast, \pi_{\xAgent[adv]}^\ast) \\
        & = r_{\xAgent[2]} + \allowbreak
         \gamma \sum_{\substack{s' \in S \\ v'_{\xAgent[ego]} \in V_{\xAgent[ego]} \\ v'_{\xAgent[adv]} \in V_{\xAgent[adv]}}} p(s',v'_{\xAgent[ego]},v'_{\xAgent[adv]}|s,v_{\xAgent[ego]},v_{\xAgent[adv]}, \sgCommonAction_{\xAgent[ego]}, \sgCommonAction_{\xAgent[adv]}) 
        \pi^{\ast}_{\xAgent[ego]}(s',v'_{\xAgent[ego]},v'_{\xAgent[adv]})  \pi^{\ast}_{\xAgent[adv]}(s',v'_{\xAgent[ego]},v'_{\xAgent[adv]}) q^{\ast}_{\xAgent[2]}(s',v'_{\xAgent[ego]},v'_{\xAgent[adv]})
         \\
         &= \sum_{\substack{s' \in S \\ v'_{\xAgent[ego]} \in V_{\xAgent[ego]} \\ v'_{\xAgent[adv]} \in V_{\xAgent[adv]}}} p(s',v'_{\xAgent[ego]},v'_{\xAgent[adv]}|s,v_{\xAgent[ego]},v_{\xAgent[adv]}, \sgCommonAction_{\xAgent[ego]}, \sgCommonAction_{\xAgent[adv]})
         \big(r_{\xAgent[2]} + \gamma \pi^{\ast}_{\xAgent[ego]}(s',v'_{\xAgent[ego]},v'_{\xAgent[adv]})  \pi^{\ast}_{\xAgent[adv]}(s',v'_{\xAgent[ego]},v'_{\xAgent[adv]}) q^{\ast}_{\xAgent[2]}(s',v'_{\xAgent[ego]},v'_{\xAgent[adv]})\big) \\
         &= \mathbb{E} [F^tq^{\ast}_{\xAgent[2]}]
    \end{aligned}
\end{equation}
As $q^{\ast}_{\xAgent[ego]} =  \mathbb{E} [F^tq^{\ast}_{\xAgent[ego]}]$ and $q^{\ast}_{\xAgent[adv]} =  \mathbb{E} [F^tq^{\ast}_{\xAgent[adv]}]$, we have $q^{\ast}=\mathbb{E}[F^tq^{\ast}]$.
\end{proof}

In Assumption 3, we introduce a global optimal point $(\Tilde{\pi}_{\xAgent[ego]}, \Tilde{\pi}_{\xAgent[adv]})$ of the stage game $(q_{\xAgent[ego]}(s,v_{\xAgent[ego]},v_{\xAgent[adv]}),q_{\xAgent[adv]}(s,v_{\xAgent[ego]},v_{\xAgent[adv]}))$, and we have $\Tilde{\pi}_{\xAgent[2]} q_{\xAgent[2]} \ge \Tilde{\pi}'_{\xAgent[2]} q_{\xAgent[2]}$ for any $\Tilde{\pi}'_{\xAgent[2]},$  $\xAgent[2] \in \{\xAgent[ego],\xAgent[adv]\}$. A global optimal point is a Nash equilibrium and all global optima are equivalent in their values.

\begin{lemma}
Let $\Tilde{\pi} = (\Tilde{\pi}_{\xAgent[ego]}, \Tilde{\pi}_{\xAgent[adv]})$ and $\Acute{\pi} =(\Acute{\pi}_{\xAgent[ego]}, \Acute{\pi}_{\xAgent[adv]})$ denote the global optimal points of the two-agent stage game $(q_{\xAgent[ego]},q_{\xAgent[adv]})$. Then, $\Tilde{\pi}q_{\xAgent[2]} = \Acute{\pi}q_{\xAgent[2]}$,  for $\xAgent[2] \in \{\xAgent[ego],\xAgent[adv] \}$.
\end{lemma}
\begin{proof}
    According to the definition of a global point,
    \begin{equation}
        \Tilde{\pi}_{\xAgent[2]} q_{\xAgent[2]} \ge \Acute{\pi}_{\xAgent[2]} q_{\xAgent[2]} 
        \text{ and }
        \Acute{\pi}_{\xAgent[2]} q_{\xAgent[2]} \ge \Tilde{\pi}_{\xAgent[2]} q_{\xAgent[2]} 
    \end{equation}
    Then, the only consistent solution is $\Tilde{\pi}_{\xAgent[2]} q_{\xAgent[2]} = \Acute{\pi}_{\xAgent[2]} q_{\xAgent[2]}$
\end{proof}

Similarly, all saddle points introduced in Assumption 3 have equal values.
\begin{lemma}
Let $\Tilde{\pi} = (\Tilde{\pi}_{\xAgent[ego]}, \Tilde{\pi}_{\xAgent[adv]}$) and $\Acute{\pi}=(\Acute{\pi}_{\xAgent[ego]}, \Acute{\pi}_{\xAgent[adv]}$) denote the saddle points of the two-agent stage game  ($q_{\xAgent[ego]},q_{\xAgent[adv]}$). Then $\Tilde{\pi}q_{\xAgent[2]} = \Acute{\pi}q_{\xAgent[2]}$, for $\xAgent[2] \in \{\xAgent[ego],\xAgent[adv] \}$.
\end{lemma}

\begin{proof}
 According to the definition of a saddle point,
 \begin{equation}
     \begin{aligned}
        & \Tilde{\pi}_{\xAgent[2]} \Tilde{\pi}_{-\xAgent[2]}q_{\xAgent[2]} \ge \Acute{\pi}_{\xAgent[2]} \Tilde{\pi}_{-\xAgent[2]}q_{\xAgent[2]} \\
        & \Acute{\pi}_{\xAgent[2]} \Acute{\pi}_{-\xAgent[2]}q_{\xAgent[2]} \le \Acute{\pi}_{\xAgent[2]} \Tilde{\pi}_{-\xAgent[2]}q_{\xAgent[2]}
     \end{aligned}
 \end{equation}
 Then, we have $\Tilde{\pi}_{\xAgent[2]} \Tilde{\pi}_{-\xAgent[2]}q_{\xAgent[2]} \ge \Acute{\pi}_{\xAgent[2]} \Acute{\pi}_{-\xAgent[2]}q_{\xAgent[2]}$. Similarly, we can have $\Acute{\pi}_{\xAgent[2]} \Acute{\pi}_{-\xAgent[2]}q_{\xAgent[2]} \ge \Tilde{\pi}_{\xAgent[2]} \Tilde{\pi}_{-\xAgent[2]}q_{\xAgent[2]}$. Then, the only consistent solution is
 \begin{equation}
     \Tilde{\pi}_{\xAgent[2]} \Tilde{\pi}_{-\xAgent[2]}q_{\xAgent[2]} = \Acute{\pi}_{\xAgent[2]} \Acute{\pi}_{-\xAgent[2]}q_{\xAgent[2]}
 \end{equation}
Thus, we have $\Tilde{\pi}q_{\xAgent[2]} = \Acute{\pi}q_{\xAgent[2]}$.
\end{proof}

In \Cref{lemma:follow}, the other condition for the convergence of $q$ to $q^{\ast}$ is that $|| F^t q - F^t q^{\ast} || \le \beta ||q-q^{\ast}|| + \lambda^t$, which requires that $F^t$ is a pseudo-contraction operator. $F^t$ maps $q \in \mathbb{Q}$ to $q^{\ast}$ when $\lambda =0$. We restrict the stage games to have a global optimal point or a saddle point at a Nash equilibrium as mentioned in Assumption 3. Under such restriction, we prove that $F^t$ is a real contraction operator, mapping every two points $q$ and $\Acute{q}$ in the space $\mathbb{Q}$ close to each other, as introduced in \Cref{lemma:map}. As a real contraction operator is a stronger version of the pseudo-contraction operator, the real contraction operator $F^t$ satisfies the condition on the contraction property in \Cref{lemma:follow}. We first define the distance between two Q-functions.
\begin{definition}
 \label{def:distance}
    For $q, \Acute{q} \in \mathbb{Q}$, the distance between Q-functions is defined as
    \begin{equation}
        ||q-\Acute{q}|| = \max_{\xAgent[2] \in \{\xAgent[ego],\xAgent[adv]\}}\max_{(s,v_{\xAgent[ego]},v_{\xAgent[adv]})} \max_{a_{\xAgent[ego]},a_{\xAgent[adv]}}||q_{\xAgent[2]}(s,v_{\xAgent[ego]},v_{\xAgent[adv]})-\Acute{q}_{\xAgent[2]}(s,v_{\xAgent[ego]},v_{\xAgent[adv]}) ||
    \end{equation}
\end{definition}
Then, we show the operator $F^t$ defined in \Cref{def:map} is a contraction mapping operator.
\begin{lemma}
    \label{lemma:map}
    $||F^t q - F^t \Acute{q}|| \le \gamma ||q-\Acute{q}||$ for all $q,\Acute{q} \in \mathbb{Q}$, where $F^t$ is defined in \Cref{def:map}.
\end{lemma}

\begin{proof}
Given $(\pi_{\xAgent[ego]},\pi_{\xAgent[adv]})$ is the Nash equilibrium strategy for the stage game $(q_{\xAgent[ego]},q_{\xAgent[adv]})$, $(\Acute{\pi}_{\xAgent[ego]},\Acute{\pi}_{\xAgent[adv]})$ is the Nash equilibrium strategy for the stage game $(\Acute{q}_{\xAgent[ego]},\Acute{q}_{\xAgent[adv]})$, we have 
\begin{equation}
    \begin{aligned}
    ||F^t q - F^t \Acute{q}|| &= \max_{\xAgent[2]}\max_{(s,v_{\xAgent[ego]},v_{\xAgent[adv]})} |\gamma \pi_{\xAgent[ego]}(s,v_{\xAgent[ego]},v_{\xAgent[adv]})\pi_{\xAgent[adv]}(s,v_{\xAgent[ego]},v_{\xAgent[adv]})q_{\xAgent[2]}(s,v_{\xAgent[ego]},v_{\xAgent[adv]}) 
    - \gamma \Acute{\pi}_{\xAgent[ego]}(s,v_{\xAgent[ego]},v_{\xAgent[adv]})\Acute{\pi}_{\xAgent[adv]}(s,v_{\xAgent[ego]},v_{\xAgent[adv]})\Acute{q}_{\xAgent[2]}(s,v_{\xAgent[ego]},v_{\xAgent[adv]}) |\\
    &= \max_{\xAgent[2]} \max_{(s,v_{\xAgent[ego]},v_{\xAgent[adv]})} \gamma |\pi_{\xAgent[ego]}(s,v_{\xAgent[ego]},v_{\xAgent[adv]})\pi_{\xAgent[adv]}(s,v_{\xAgent[ego]},v_{\xAgent[adv]})q_{\xAgent[2]}(s,v_{\xAgent[ego]},v_{\xAgent[adv]}) 
    - \Acute{\pi}_{\xAgent[ego]}(s,v_{\xAgent[ego]},v_{\xAgent[adv]})\Acute{\pi}_{\xAgent[adv]}(s,v_{\xAgent[ego]},v_{\xAgent[adv]})\Acute{q}_{\xAgent[2]}(s,v_{\xAgent[ego]},v_{\xAgent[adv]})|
    \end{aligned}
\end{equation}
To simplify the notation, we omit the inputs of $\pi \text{ and } \Acute{\pi}$.
The next step is to show 
\begin{equation}
    |\pi_{\xAgent[ego]}\pi_{\xAgent[adv]}q_{\xAgent[2]}(s,v_{\xAgent[ego]},v_{\xAgent[adv]}) - \Acute{\pi}_{\xAgent[ego]}\Acute{\pi}_{\xAgent[adv]}\Acute{q}_{\xAgent[2]}(s,v_{\xAgent[ego]},v_{\xAgent[adv]})| \le
    ||q(s,v_{\xAgent[ego]},v_{\xAgent[adv]})-\Acute{q}(s,v_{\xAgent[ego]},v_{\xAgent[adv]})||
\end{equation}
1. Under condition 1 in Assumption 3:

Both $(\pi_{\xAgent[ego]},\pi_{\xAgent[adv]})$ and $(\Acute{\pi}_{\xAgent[ego]},\Acute{\pi}_{\xAgent[adv]})$ are global optimal points.

If $\pi_{\xAgent[ego]}\pi_{\xAgent[adv]}q_{\xAgent[2]}(s, v_{\xAgent[ego]},v_{\xAgent[adv]}) \ge \Acute{\pi}_{\xAgent[ego]}\Acute{\pi}_{\xAgent[adv]}\Acute{q}_{\xAgent[2]}(s, v_{\xAgent[ego]},v_{\xAgent[adv]})$, we have
\begin{equation}
    \begin{aligned}
    & \pi_{\xAgent[ego]}\pi_{\xAgent[adv]}q_{\xAgent[2]}(s,v_{\xAgent[ego]},v_{\xAgent[adv]}) - \Acute{\pi}_{\xAgent[ego]}\Acute{\pi}_{\xAgent[adv]}\Acute{q}_{\xAgent[2]}(s,v_{\xAgent[ego]},v_{\xAgent[adv]}) \\
    \le & \pi_{\xAgent[ego]}\pi_{\xAgent[adv]}q_{\xAgent[2]}(s,v_{\xAgent[ego]},v_{\xAgent[adv]}) - \pi_{\xAgent[ego]}\pi_{\xAgent[adv]}\Acute{q}_{\xAgent[2]}(s,v_{\xAgent[ego]},v_{\xAgent[adv]}) \\
    = & \sum_{a_{\xAgent[ego]},a_{\xAgent[adv]}} \pi_{\xAgent[ego]}(a_{\xAgent[ego]})\pi_{\xAgent[adv]}(a_{\xAgent[adv]}) (q_{\xAgent[2]}(s, v_{\xAgent[ego]},v_{\xAgent[adv]},a_{\xAgent[ego]},a_{\xAgent[adv]})  - \Acute{q}_{\xAgent[2]}(s, v_{\xAgent[ego]},v_{\xAgent[adv]},a_{\xAgent[ego]},a_{\xAgent[adv]}))\\
    \le & \sum_{a_{\xAgent[ego]},a_{\xAgent[adv]}} \pi_{\xAgent[ego]}(a_{\xAgent[ego]})\pi_{\xAgent[adv]}(a_{\xAgent[adv]}) ||q_{\xAgent[2]}(s, v_{\xAgent[ego]},v_{\xAgent[adv]}) - \Acute{q}_{\xAgent[2]}(s, v_{\xAgent[ego]},v_{\xAgent[adv]}) ||\\
    \le & || q_{\xAgent[2]}(s, v_{\xAgent[ego]},v_{\xAgent[adv]}) - \Acute{q}_{\xAgent[2]}(s, v_{\xAgent[ego]},v_{\xAgent[adv]})||
    \end{aligned}
\end{equation}
The first inequality derives from the property of the global point that $\Acute{\pi}_{\xAgent[ego]}\Acute{\pi}_{\xAgent[adv]}\Acute{q}_{\xAgent[2]}(s,v_{\xAgent[ego]},v_{\xAgent[adv]}) \ge \pi_{\xAgent[ego]}\pi_{\xAgent[adv]}\Acute{q}_{\xAgent[2]}(s,v_{\xAgent[ego]},v_{\xAgent[adv]})$. 
The second inequality is based on $||q_{\xAgent[2]}(s, v_{\xAgent[ego]},v_{\xAgent[adv]}) - \Acute{q}_{\xAgent[2]}(s, v_{\xAgent[ego]},v_{\xAgent[adv]}) || = \max_{a_{\xAgent[ego]},a_{\xAgent[adv]}} |q_{\xAgent[2]}(s, v_{\xAgent[ego]},v_{\xAgent[adv]},a_{\xAgent[ego]},a_{\xAgent[adv]}) - \Acute{q}_{\xAgent[2]}(s, v_{\xAgent[ego]},v_{\xAgent[adv]},a_{\xAgent[ego]},a_{\xAgent[adv]}) |$.

If $\pi_{\xAgent[ego]}\pi_{\xAgent[adv]}q_{\xAgent[2]}(s, v_{\xAgent[ego]},v_{\xAgent[adv]}) \le \Acute{\pi}_{\xAgent[ego]}\Acute{\pi}_{\xAgent[adv]}\Acute{q}_{\xAgent[2]}(s, v_{\xAgent[ego]},v_{\xAgent[adv]})$, we have
\begin{equation}
    \begin{aligned}
        & \Acute{\pi}_{\xAgent[ego]}\Acute{\pi}_{\xAgent[adv]}\Acute{q}_{\xAgent[2]}(s, v_{\xAgent[ego]},v_{\xAgent[adv]}) - \pi_{\xAgent[ego]}\pi_{\xAgent[adv]}q_{\xAgent[2]}(s, v_{\xAgent[ego]},v_{\xAgent[adv]}) 
        \le \Acute{\pi}_{\xAgent[ego]}\Acute{\pi}_{\xAgent[adv]}\Acute{q}_{\xAgent[2]}(s, v_{\xAgent[ego]},v_{\xAgent[adv]}) - \Acute{\pi}_{\xAgent[ego]}\Acute{\pi}_{\xAgent[adv]}q_{\xAgent[2]}(s, v_{\xAgent[ego]},v_{\xAgent[adv]})
    \end{aligned}
\end{equation}
and the rest of the proof is similar to the above.

2. Under condition 2 in Assumption 3:

Both $(\pi_{\xAgent[ego]},\pi_{\xAgent[adv]})$ and $(\Acute{\pi}_{\xAgent[ego]},\Acute{\pi}_{\xAgent[adv]})$ are saddle points. 

If $\pi_{\xAgent[ego]}\pi_{\xAgent[adv]}q_{\xAgent[2]}(s, v_{\xAgent[ego]},v_{\xAgent[adv]}) \ge \Acute{\pi}_{\xAgent[ego]}\Acute{\pi}_{\xAgent[adv]}\Acute{q}_{\xAgent[2]}(s, v_{\xAgent[ego]},v_{\xAgent[adv]})$, we have 
\begin{equation}
    \begin{aligned}
        & \pi_{\xAgent[ego]}\pi_{\xAgent[adv]}q_{\xAgent[2]}(s, v_{\xAgent[ego]},v_{\xAgent[adv]}) -  \Acute{\pi}_{\xAgent[ego]}\Acute{\pi}_{\xAgent[adv]}\Acute{q}_{\xAgent[2]}(s, v_{\xAgent[ego]},v_{\xAgent[adv]}) \\
        & \le \pi_{\xAgent[ego]}\pi_{\xAgent[adv]}q_{\xAgent[2]}(s, v_{\xAgent[ego]},v_{\xAgent[adv]}) - \pi_{\xAgent[ego]}\Acute{\pi}_{\xAgent[adv]}\Acute{q}_{\xAgent[2]}(s, v_{\xAgent[ego]},v_{\xAgent[adv]}) \\
        & \le 
        \pi_{\xAgent[ego]}\Acute{\pi}_{\xAgent[adv]}q_{\xAgent[2]}(s, v_{\xAgent[ego]},v_{\xAgent[adv]}) - \pi_{\xAgent[ego]}\Acute{\pi}_{\xAgent[adv]}\Acute{q}_{\xAgent[2]}(s, v_{\xAgent[ego]},v_{\xAgent[adv]}) \\
        & \le||q_{\xAgent[2]}(s, v_{\xAgent[ego]},v_{\xAgent[adv]}) - \Acute{q}_{\xAgent[2]}(s, v_{\xAgent[ego]},v_{\xAgent[adv]})||
    \end{aligned}
\end{equation}
The inequalities are based on the properties of saddle points, which are mentioned in Assumption 3.

If $\pi_{\xAgent[ego]}\pi_{\xAgent[adv]}q_{\xAgent[2]}(s, v_{\xAgent[ego]},v_{\xAgent[adv]}) \le \Acute{\pi}_{\xAgent[ego]}\Acute{\pi}_{\xAgent[adv]}\Acute{q}_{\xAgent[2]}(s, v_{\xAgent[ego]},v_{\xAgent[adv]})$, the proof is similar to the above.

Thus, 
\begin{equation}
\begin{aligned}
    ||F^t q - F^t \Acute{q}|| &=
    \max_{\xAgent[2]} \max_{(s,v_{\xAgent[ego]},v_{\xAgent[adv]})} \gamma |\pi_{\xAgent[ego]}(s,v_{\xAgent[ego]},v_{\xAgent[adv]})\pi_{\xAgent[adv]}(s,v_{\xAgent[ego]},v_{\xAgent[adv]})q_{\xAgent[2]}(s,v_{\xAgent[ego]},v_{\xAgent[adv]}) 
    - \Acute{\pi}_{\xAgent[ego]}(s,v_{\xAgent[ego]},v_{\xAgent[adv]})\Acute{\pi}_{\xAgent[adv]}(s,v_{\xAgent[ego]},v_{\xAgent[adv]})\Acute{q}_{\xAgent[2]}(s,v_{\xAgent[ego]},v_{\xAgent[adv]})| \\
    &\le \max_{\xAgent[2]} \max_{(s,v_{\xAgent[ego]},v_{\xAgent[adv]})} \gamma 
    ||q_{\xAgent[2]}(s, v_{\xAgent[ego]},v_{\xAgent[adv]}) - \Acute{q}_{\xAgent[2]}(s, v_{\xAgent[ego]},v_{\xAgent[adv]}) || \\
    & \le \gamma||q - \Acute{q} ||
\end{aligned}    
\end{equation}
\end{proof}

Based on \Cref{lemma:map} and \Cref{lemma:exp}, the operator $F^t$ defined for two-agent stochastic game in \Cref{def:map} satisfies the condition for convergence in \Cref{lemma1:follow}. Moreover, the updating function of Q-functions in {\algoName} uses the same updating method in \Cref{lemma:follow}. Thus, we apply \Cref{lemma:follow} to {\algoName} and achieve Theorem 1 that Q-functions in {\algoName} converge to Q-functions at a Nash equilibrium under certain conditions.


\section{Effect of $\epsilon$ on {\algoName}}

When using the $\epsilon$-greedy policy, $\epsilon$ determines the probability of taking a random action at each step during training, which affects the balance between exploration and exploitation and
would further affect the learning speed~\cite{van2016deep,lin2017explore}.
To study the effect of $\epsilon$ on {\algoName}, we compare the performance of {\algoName} with varying $\epsilon$ values to learn the task in Case Study I, including $\epsilon=0.05$, $\epsilon=0.25$ (which is adopted in Section 6), $\epsilon = 0.90$, and $\epsilon = 0.90$ decaying with $\epsilon\times 0.9986$ until $\epsilon$ reaches 0.05 (which is adopted in ~\cite{zamora2016extending}). 

\begin{figure}[hbt!]
        \centering
        \includegraphics[width=0.7\linewidth]{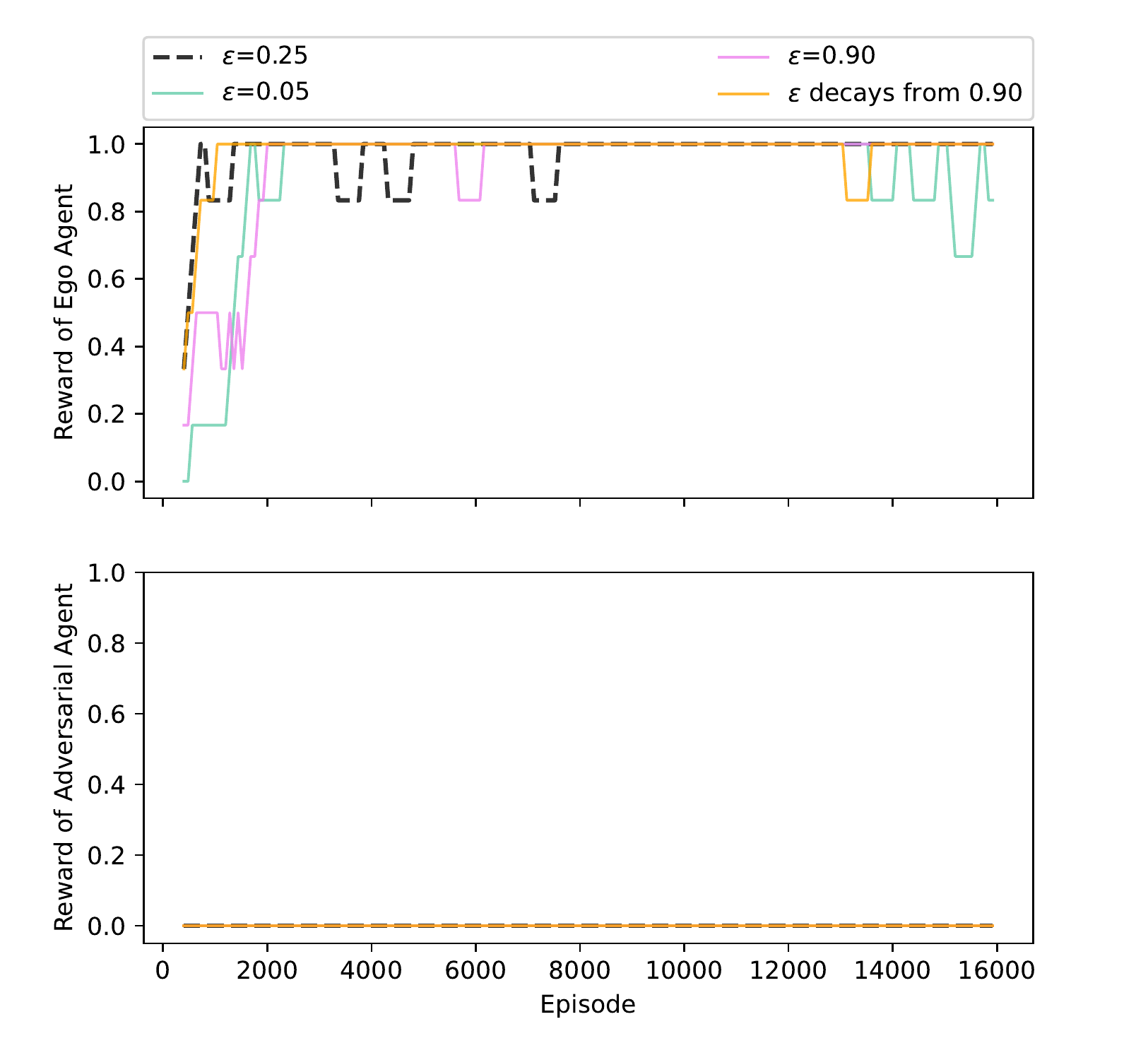}
        \caption{Cumulative reward comparison for learning the task introduced in Case Study I using QRM-SG with varying $\epsilon$. Smoothed plot with a rolling window of size 6. The adversarial agent receives a cumulative reward of 0 with all $\epsilon$ values.}
        \label{fig:epsilon}
\end{figure}

We evaluate the performance of {\algoName} with different $\epsilon$ every 80 episodes during training, and \Cref{fig:epsilon} plots the cumulative rewards of each agent during the evaluations. We note that $\epsilon$ is set to 0 during the evaluations. When using {\algoName} with $\epsilon$ of 0.05 or 0.90, the ego agent starts to complete the task after around 2000 episodes, significantly more training episodes compared to {\algoName} with $\epsilon$ of 0.25 or decaying $\epsilon$. 
The reason is that $\epsilon$ with a large value focuses on exploring the environment to collect data about different actions and their outcomes, whereas $\epsilon$ with a small value focuses on actions that have previously resulted in higher rewards. 
In this case study, $\epsilon$ of 0.25 strikes a balance between exploration and exploitation, allowing the agent to efficiently learn the complex task. 

\section{Evaluation of {\algoName} on a 12$\times$12 Grid World}

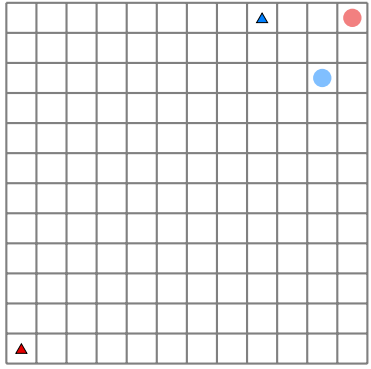
\begin{figure}
		\centering
		\input{figs/tasks-map2-large.tikz}
		\caption{
		    The 12$\times$12 grid world for scalability analysis of QRM-SG. Locations \protect\location[EgoHome] and \protect\location[AdvHome]  denote the power bases of the ego agent and adversarial agent, respectively. \protect\agent[Ego] and \protect\agent[Adv] indicate the starting locations of the ego agent and adversarial agent, respectively.
		    } \label{fig:grid_map_large}
\end{figure}

\begin{figure}[hbt!]
        \centering
        \includegraphics[width=0.6\linewidth]{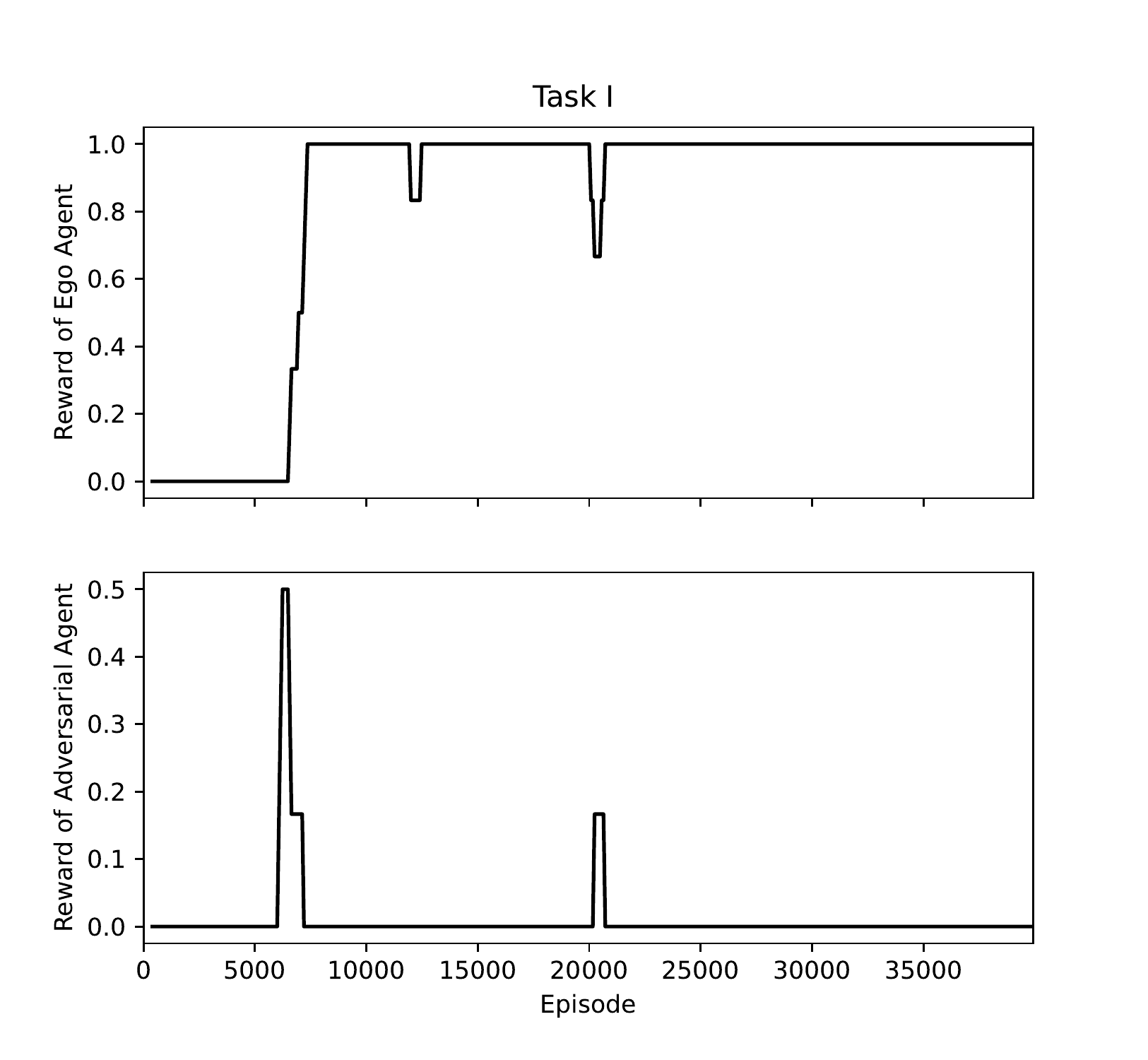}
        \caption{Cumulative reward of learning the task introduced in Case Study I in the 12$\times$12 grid world using QRM-SG. Smoothed plot with a rolling window of size 6.}
        \label{fig:large}
\end{figure}

To further evaluate the performance of {\algoName}, we generalize {\algoName} to learn the task introduced in Case Study I in a 12$\times$12 grid world. The grid world size aligns with the case studies in other existing research papers, such as the 9$\times$12 office world scenario in~\cite{DBLP:conf/icml/IcarteKVM18} and 15$\times$15 grid world in~\cite{lopes2009active}. As shown in \Cref{fig:grid_map_large}, the adversarial agent is far away from its power base, thus it is expected to fail to complete its own task, whereas the ego agent is expected to complete its own task. \Cref{fig:large} shows the learning process of each agent. 
Similar to the case studies in Section 6, every 80 episodes, we stop learning, test the algorithm's performance, and
save the cumulative rewards of each agent. 
The ego agent starts to complete the task after around 7000 episodes and {\algoName} finds the Nash equilibrium in around 21000 episodes.

%% file: figs/tasks-map2-large.tikz
\begin{tikzpicture}[
    scale=0.8,
    thick,
    every node/.append style={transform shape},
]

\begin{scope}[
    xscale=.5,
    yscale=.5, 
    every node/.append style={scale=2}, 
]

\draw[step=1,color=gray,shift={(-.5,-.5)}] (0,0) grid +(12,12);

\node at (10,9) {\Large\location[EgoHome]};
\node at (11,11) {\Large\location[AdvHome]};
\node at (8,11) {\agent[Ego]};
\node at (0,0) {\agent[Adv]}; 

\end{scope}

\end{tikzpicture}